\documentclass[11pt]{article}
\usepackage[margin=1in]{geometry}
\usepackage[T1]{fontenc}
\usepackage{lmodern}
\usepackage{microtype}
\usepackage{setspace}
\usepackage{graphicx}
\usepackage{booktabs}
\usepackage{multirow}
\usepackage{array}
\usepackage{tabularx}
\usepackage{longtable}
\usepackage{amsmath,amssymb,mathtools,bm}
\usepackage{siunitx}
\usepackage{xcolor}
\usepackage{caption}
\usepackage{subcaption}
\usepackage{float}
\usepackage{placeins}
\usepackage{enumitem}
\usepackage[round,authoryear]{natbib}
\usepackage{multibib}
\newcites{app}{Appendix References}
\usepackage[colorlinks=true,linkcolor=blue!55!black,citecolor=blue!55!black,urlcolor=blue!55!black]{hyperref}
\hypersetup{pdftitle={The Global Mediation Workspace: A Control-Theoretic Formulation of Global Workspace Theory},pdfauthor={Ryota Kanai},pdfsubject={A control-theoretic formulation of global workspace theory},pdfkeywords={Global Mediation Workspace, global neuronal workspace, controllability, observability, boundary Hankel operator, nonlinear dynamical systems, macaque ECoG, anesthesia, mode alignment, global access}}
\usepackage[nameinlink,capitalize,noabbrev]{cleveref}
\graphicspath{{figures_v14/}{figures_v13/}{figures_v12/}{figures_v3/}{figures_v2/}{figures_nonlinear/}{figures_macaque_corrected/}{figures_macaque_final/}{figures_macaque_officialmap/}}
\newcommand{\Reach}{\mathcal{R}}
\newcommand{\Observe}{\mathcal{O}}
\newcommand{\Hankel}{\mathcal{H}}
\newcommand{\Wc}{W_{\mathrm c}}
\newcommand{\Wo}{W_{\mathrm o}}
\newcommand{\tr}{\operatorname{tr}}
\newcommand{\im}{\operatorname{im}}
\newcommand{\rank}{\operatorname{rank}}
\newcommand{\Deff}{D_{\mathrm{eff}}}
\newcommand{\DeffL}{D_{\mathrm{eff},L}}
\newcommand{\WMI}{\operatorname{WMI}}
\newcommand{\Cspec}{C_{\mathrm{spec}}}
\newcommand{\CspecL}{C_{\mathrm{spec},L}}
\newcommand{\Aspec}{A_{\mathrm{spec}}}
\newcommand{\AspecL}{A_{\mathrm{spec},L}}
\newcommand{\Ctot}{C_{\mathrm{tot}}}
\newcommand{\AF}{A_{\mathrm F}}
\newcommand{\Oorg}{O_{\mathrm{org}}}
\newcommand{\Gsig}{\mathfrak{G}}
\newcommand{\normnuc}[1]{\left\lVert #1\right\rVert_*}
\newcommand{\normF}[1]{\left\lVert #1\right\rVert_{\mathrm F}}
\newcolumntype{Y}{>{\raggedright\arraybackslash}X}
\newcolumntype{C}{>{\centering\arraybackslash}X}

\title{\textbf{A Control-Theoretic Formulation of \\ Global Workspace Theory}
}
\author{Ryota Kanai\\
\small Araya Inc., Tokyo, Japan}

\begin{document}
\maketitle

\begin{abstract}
Global workspace theory explains conscious access in terms of the availability of selected information to specialized systems, but it does not provide a formal criterion for identifying the mechanism that enables this access. We introduce the \emph{Global Mediation Workspace} (GMW), a control-theoretic formulation in which a candidate subnetwork is treated as an open system embedded in the remainder of the network. Finite-horizon boundary reachability characterizes how activity in the remainder can drive the candidate, observability characterizes how candidate states can affect the remainder, and a boundary Hankel operator identifies the internal modes that connect these two directions. The resulting signature separates potential mediation capacity, input--output alignment, effective dimensionality, and routed source--target breadth. No single component is stipulated to be the defining property of consciousness. Which components, or combinations of components, are necessary for conscious access remains an empirical question.

We validated the framework using synthetic benchmarks designed to distinguish a planted differentiated mediator from dense hubs, one-sided receivers or broadcasters, and a split read/write aggregate with no common internal route. The GMW signature separated the planted mediator from these alternative network structures. We also developed a nonlinear extension in which mediation is characterized by trajectory-conditioned differential operators, finite-amplitude response profiles, and state-dependent coalitions that instantiate the GMW.

As a preliminary assessment of applicability, we then applied the framework to geometry-audited, nonoverlapping bipolar ECoG recordings from four macaques across 11 ketamine--medetomidine experiments. This analysis asks whether the signature can be estimated from real neural recordings and which estimation problems arise in practice, rather than testing global workspace theory itself. Deep anesthesia increased short-lag predictability, realized mediation strength, and potential capacity, while reducing input--output alignment. Reductions in effective dimensionality, routed breadth, and gain-free organization were most apparent at the largest candidate size and varied across animals. Repeatedly selected awake candidate sites were distributed across frontal, parietal, temporal, and other association-related cortical regions. The application shows that the GMW signature can separate changes in dynamical gain from changes in mode organization, while exposing methodological constraints on montage construction, candidate selection, and temporal structure. Taken together, these results establish the GMW as a computational framework for identifying and characterizing subnetworks that receive activity from distributed systems, transform it through internal modes, and return differentiated effects to the wider network.
\end{abstract}


\section{Introduction}

Global workspace theory addresses a central organizational problem: how can a system composed of specialized processes coordinate flexible cognition around a common content? In the original cognitive formulation, information admitted to a limited-capacity workspace becomes available to perception, memory, evaluation, reasoning, and action systems that otherwise operate with partly distinct representations and goals \citep{Baars1988}. The global neuronal workspace account connects this functional transition to recurrent amplification and widespread reciprocal interactions, producing an ignition-like shift from local processing to broad availability \citep{Dehaene1998,DehaeneNaccache2001,DehaeneChangeux2011,Mashour2020}. In this account, the workspace is a dynamical organization and is not identified with a single anatomical location.

The theory nevertheless leaves an operational question unresolved. Given a network dynamical model, which subnetwork performs the workspace operation? A dense hub may contact many regions while transmitting only one redundant mode. A receiver may encode activity elsewhere without influencing it, and a broadcaster may exert broad effects while receiving little. More subtly, a set can contain excellent receivers and excellent broadcasters with no internal route connecting what is received to what is sent. Separate input and output capacities are insufficient.

Several mathematical approaches address parts of the workspace idea. The Global Latent Workspace emphasizes bidirectional translation between heterogeneous representational spaces through a shared latent format \citep{VanRullenKanai2021}. Partial information decomposition and Integrated Information Decomposition distinguish redundant from synergistic information and have motivated synergistic cores, gateways, and broadcasters in neuroimaging data \citep{WilliamsBeer2010,Mediano2025,Luppi2022,Luppi2024}. These approaches clarify representational compatibility and multivariate information structure. They do not by themselves specify the directional, finite-time mechanism through which a designated subsystem is driven, transforms its state, and becomes consequential for other systems.

Control theory supplies an input--state--output language for this problem \citep{Kailath1980,Moore1981}. In a controlled system, controllability asks which internal state directions can be reached by inputs delivered over time and how strongly those directions can be excited. In observational neural data, the same construction is interpreted more narrowly as reachability from a declared boundary because the boundary variables are not independently manipulated. Observability asks which internal-state differences can be recovered from their effects on outputs. High degree does not guarantee either property: many connections can collapse onto a narrow state direction, and strong output can fail to distinguish internal states. Both quantities depend on the dynamics, the declared input and output channels, the state metric, and the time horizon.

\begin{figure}[t]
    \centering
    \includegraphics[width=\textwidth]{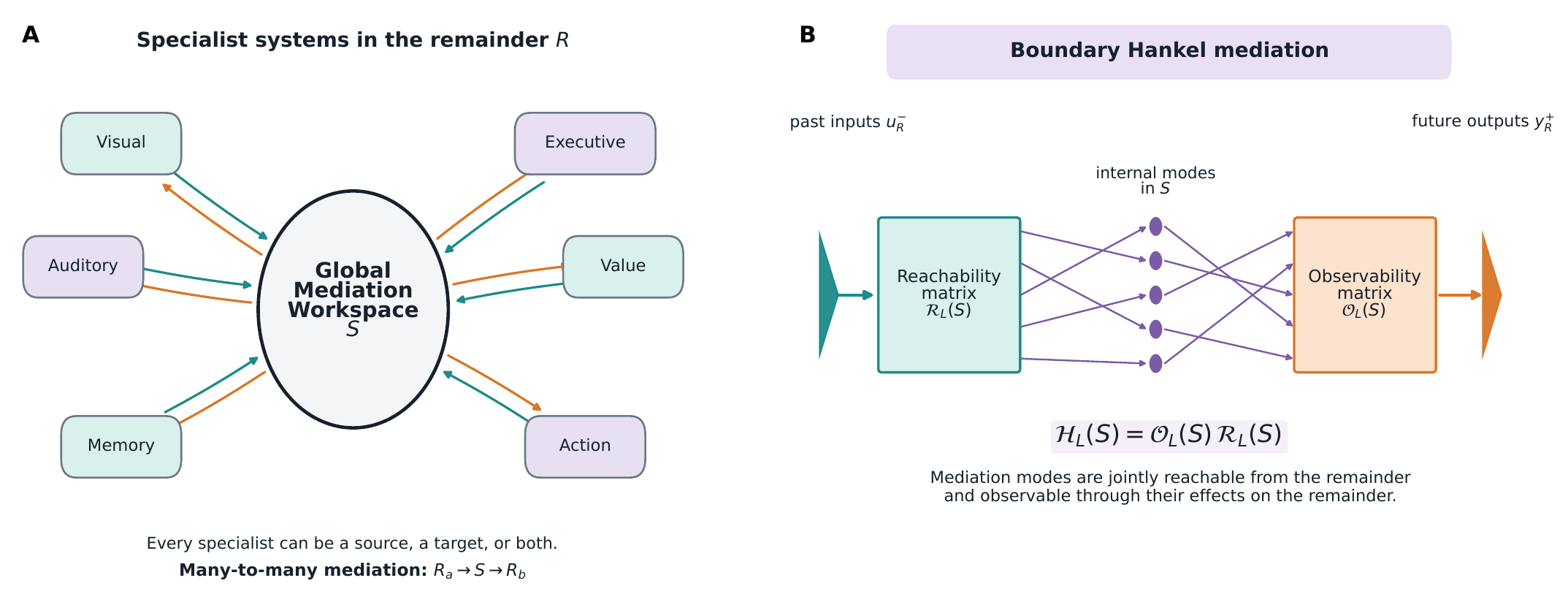}
    \caption{\textbf{The Global Mediation Workspace and its boundary operator.} (A) The candidate $S$ is embedded in the remainder $R$, and every specialist system can act as a source, a target, or both. (B) The reachability matrix $\mathcal R_L(S)$ maps past boundary inputs into candidate states, and the observability matrix $\mathcal O_L(S)$ maps those states to future boundary outputs. Their product $\mathcal H_L(S)=\mathcal O_L(S)\mathcal R_L(S)$ identifies internal modes that are jointly reachable from and observable in the remainder.}
    \label{fig:gmw-concept}
\end{figure}

For a global workspace, the same collection of specialist systems occupies both sides of the relation: the remainder of the network must be able to drive a candidate, and the candidate must return consequential activity to that same remainder. We call a candidate that performs this operation a \emph{Global Mediation Workspace} (GMW). Its defining function is many-to-many mediation. Any specialist subsystem may supply activity to the candidate, receive activity from it, or do both. The relevant operation is the map $R\rightarrow S\rightarrow R$, not a fixed chain with privileged input and output modules (\cref{fig:gmw-concept}A).

At the operator level, past boundary inputs are mapped into candidate states by a finite-horizon reachability matrix, and those states are mapped to future boundary outputs by an observability matrix (\cref{fig:gmw-concept}B). Their product is the boundary Hankel operator. Its singular modes identify candidate-state directions that are jointly reachable from the remainder and observable through their effects on that remainder. We use this operator to relate the system-level idea of global availability to a measurable read--transform--write process.

The singular spectrum also provides measures of capacity and scale. Total spectral strength measures realized read--transform--write gain, while effective dimensionality separates a repertoire of independent routes from a single repeated bottleneck. Comparing the controllability and observability spectra with the realized joint spectrum further distinguishes available capacity from the alignment that makes it usable.

The framework developed below treats this spectrum as the primary dynamical object. A four-component signature summarizes capacity, alignment, differentiated mode count, and routed source--target breadth. The Workspace Mediation Index (WMI) is retained as a pragmatic scalar for fixed-size candidate search, not as the definition of the GMW and not as a scalar consciousness measure. Which signature components are necessary for conscious access, and how they depend on task, state, and measurement scale, must be determined empirically.

After establishing the linear theory, we test it in synthetic networks constructed to include confounding motifs and extend it to nonlinear, trajectory-dependent systems. We then apply the formulation to high-density macaque ECoG during wakefulness, deep ketamine--medetomidine anesthesia, and recovery. In this setting, anesthesia can increase slow-wave gain and short-lag predictability while differentiated global organization declines. The analysis tests whether the GMW signature separates interaction magnitude from its mode organization.

\section{Problem formulation and inferential scope}
\label{sec:problem}

Let a directed discrete-time dynamical network be locally represented by
\begin{equation}
    x_{t+1}=Ax_t,
    \label{eq:autonomous}
\end{equation}
where $x_t\in\mathbb R^n$ and $A_{ij}$ is the linear predictive influence from source node $j$ to target node $i$. Depending on the application, $A$ may be a local Jacobian, an effective-connectivity estimate, a latent-state transition, or a known generative operator. It should not be identified automatically with anatomical adjacency \citep{Friston2011,Ljung1999}.

For a candidate set $S\subset V$ with $k=|S|$, let $R=V\setminus S$ and partition
\begin{equation}
\begin{bmatrix}x_S(t+1)\\x_R(t+1)\end{bmatrix}
=
\begin{bmatrix}
A_{SS}&A_{SR}\\
A_{RS}&A_{RR}
\end{bmatrix}
\begin{bmatrix}x_S(t)\\x_R(t)\end{bmatrix}.
\label{eq:partition}
\end{equation}
We write
\begin{equation}
    A_S=A_{SS},\qquad B_S=A_{SR},\qquad C_S=A_{RS},
\end{equation}
and view the candidate as the open subsystem
\begin{equation}
    z_{t+1}=A_Sz_t+B_Su_t,\qquad y_t=C_Sz_t.
    \label{eq:open}
\end{equation}
The remainder supplies boundary inputs $u_t$ and receives boundary outputs $y_t$. If $u_t$ is an independently specified intervention, the receive-side construction has the usual controllability interpretation. In the observational neural analysis below, $u_t$ denotes modeled activity in the remainder entering through $B_S$. The corresponding Gramian is interpreted as boundary reachability under the fitted dynamics. It does not show that an experimenter can control the boundary coordinates independently.

A GMW candidate should have enough receive and send capacity to support substantial transfer, alignment between the internal directions reached from the boundary and those that affect the boundary, several nonredundant mediation modes, and broad routing among relevant specialist systems. Minimality and canonicality are separate questions: the framework evaluates a declared candidate, state representation, temporal grain, and system boundary. It does not by itself prove that those choices are intrinsic to the system. The main notation is summarized in \cref{app:notation}.

\section{Finite-horizon boundary mediation}
\label{sec:linear-mediation}

For horizon $L\ge1$, define the finite-horizon reachability and observability matrices
\begin{align}
    \Reach_L(S)&=\left[B_S,\ A_SB_S,\ \ldots,\ A_S^{L-1}B_S\right],
    \label{eq:reach}\\
    \Observe_L(S)&=
    \begin{bmatrix}
    C_S\\ C_SA_S\\ \vdots\\ C_SA_S^{L-1}
    \end{bmatrix}.
    \label{eq:observe}
\end{align}
Their product
\begin{equation}
    \boxed{\Hankel_L(S)=\Observe_L(S)\Reach_L(S)}
    \label{eq:hankel}
\end{equation}
is the finite-horizon boundary Hankel matrix. With zero-based block indices $i,j\in\{0,\ldots,L-1\}$,
\begin{equation}
    [\Hankel_L(S)]_{ij}=C_SA_S^{i+j}B_S.
    \label{eq:markov}
\end{equation}
Each block aggregates signed, gain-weighted paths that enter $S$, propagate within it, and return to the boundary. Paths that bypass $S$ through $A_{RR}$ are excluded because the question is whether the candidate mediates the transfer. Under the update convention in \cref{eq:open}, block $(i,j)$ links the past input $u_{t-1-j}$ to the future output $y_{t+i}$, an input--output separation of $(i+j+1)\Delta t$. For the shifted operator with $q$ required internal transitions defined in \cref{sec:shifted-hankel}, the separation is $(i+j+q+1)\Delta t$. Thus each side contains $L$ samples, while the represented input--output separations range from $(q+1)\Delta t$ to $(2L+q-1)\Delta t$.

The associated finite-horizon Gramians are
\begin{align}
    \Wc(S;L)&=\Reach_L\Reach_L^{\top}
    =\sum_{\tau=0}^{L-1}A_S^{\tau}B_SB_S^{\top}(A_S^{\top})^{\tau},
    \label{eq:wc}\\
    \Wo(S;L)&=\Observe_L^{\top}\Observe_L
    =\sum_{\tau=0}^{L-1}(A_S^{\top})^{\tau}C_S^{\top}C_SA_S^{\tau}.
    \label{eq:wo}
\end{align}
Standard balanced-system identities imply that the nonzero singular values of $\Hankel_L$ are
\begin{equation}
    \eta_i(S;L)=\sigma_i(\Hankel_L)
    =\sqrt{\lambda_i(\Wo\Wc)}
    =\sqrt{\lambda_i(\Wo^{1/2}\Wc\Wo^{1/2})}.
    \label{eq:eta}
\end{equation}
These are the finite-horizon mediation singular values. They quantify the internal directions that are simultaneously reachable from the remainder and observable through their effects on it \citep{Moore1981,Kurschner2018}. The equivalence, internal similarity invariance, shifted operators that require a minimum internal path length, and a path-constrained cross-node variant are collected in \cref{app:linear-identities}.

\section{Capacity and alignment of mediation modes}
\label{sec:capacity-alignment}

Let
\begin{equation}
    \Wc=U_c\Lambda_cU_c^\top,\qquad
    \Wo=U_o\Lambda_oU_o^\top,
    \label{eq:eigendecomp}
\end{equation}
with eigenvalues ordered nonincreasingly, and define the relative mode basis
\begin{equation}
    M=U_o^\top U_c.
    \label{eq:M}
\end{equation}
The mediation singular values can be written as
\begin{equation}
    \boxed{\eta_i=\sigma_i\!\left(\Lambda_o^{1/2}M\Lambda_c^{1/2}\right).}
    \label{eq:crossfactor}
\end{equation}
The cross-basis factorization separates these two contributions. The eigenvalues of $\Wc$ and $\Wo$ describe receive and send capacity separately. Their energetic directions produce realized mediation only to the extent that the two mode bases are aligned. Principal-angle diagnostics and derivations are given in \cref{app:alignment-details} \citep{Moore1981,BjorckGolub1973}.

To separate potential capacity from realized alignment, define
\begin{equation}
    \Cspec=\sum_{i=1}^{k}\sqrt{\lambda_i(\Wc)\lambda_i(\Wo)}
    \label{eq:cspec}
\end{equation}
and
\begin{equation}
    \Aspec=\frac{Q_L}{\Cspec},\qquad
    Q_L=\normnuc{\Hankel_L}=\sum_i\eta_i,
    \label{eq:aspec}
\end{equation}
with $\Aspec=0$ when $\Cspec=0$. The resulting decomposition is
\begin{equation}
    \boxed{Q_L=\Cspec\Aspec,\qquad 0\leq\Aspec\leq1.}
    \label{eq:decomp}
\end{equation}
The upper bound follows from a standard singular-value rearrangement inequality; a derivation is included in \cref{app:linear-identities}.

$\Cspec$ is the largest mediation strength compatible with the two Gramian spectra when their modes are optimally paired. $\Aspec$ is the fraction of that envelope realized by the actual mode alignment. A candidate can have substantial $\Cspec$ but small $\Aspec$ when its reachable and observable directions are mismatched; the split read/write aggregate in the synthetic benchmark illustrates this case. A dense hub can show the opposite limitation: near-perfect alignment concentrated in one mode. Capacity and alignment should be interpreted together with spectral dimensionality.

\section{The GMW signature and fixed-size search}
\label{sec:signature}

Let $\bm\eta=(\eta_1,\ldots,\eta_k)$ be ordered from largest to smallest. For $Q_L>0$, define
\begin{equation}
    p_i=\frac{\eta_i}{Q_L},\qquad
    \DeffL(S)=\exp\!\left[-\sum_{i:p_i>0}p_i\log p_i\right],
    \label{eq:erank}
\end{equation}
and set $\DeffL=0$ when $Q_L=0$. The value is one for a rank-one mediator and approaches $k$ for a flat spectrum.

To measure routed breadth, partition specialist nodes in $R$ into modules $M_1,\ldots,M_m$. For source module $a$ and target module $b$, let $\Reach_{a,L}$ and $\Observe_{b,L}$ be the corresponding boundary matrices. The routed operator and its energy are
\begin{equation}
    \Hankel_L^{b\leftarrow a}=\Observe_{b,L}\Reach_{a,L},\qquad
    \gamma_{ba}=\normF{\Hankel_L^{b\leftarrow a}}^2
    =\tr(\Wo{}_b\Wc{}_a).
    \label{eq:gamma}
\end{equation}
After normalizing $\pi_{ba}=\gamma_{ba}/\sum_{c,d}\gamma_{cd}$, define
\begin{equation}
    G_{\mathrm{pair},L}(S)=
    \frac{\exp[-\sum_{a,b:\pi_{ba}>0}\pi_{ba}\log\pi_{ba}]}{m^2},
    \label{eq:gpair}
\end{equation}
with value zero when the total routed energy is zero. This quantity measures breadth of mediated source--target transformations, not diversity of incident edges alone.

When the horizon is shown explicitly, we write $\CspecL$ and $\AspecL$ for the quantities in \cref{eq:cspec,eq:aspec}. The primary descriptive object is the four-component GMW signature
\begin{equation}
    \boxed{\Gsig_L(S)=
    \left(\CspecL,\ \AspecL,\ \DeffL/|S|,\ G_{\mathrm{pair},L}\right).}
    \label{eq:signature}
\end{equation}
At a fixed candidate size, a GMW candidate is a set whose signature indicates substantial aligned capacity, differentiated mediation, and broad routed access relative to appropriately matched candidates. No single scalar defines the GMW. Which components of \cref{eq:signature} are necessary for conscious access is an empirical question, not a stipulation of the framework.

For fixed-size search, we use the pragmatic scalarization
\begin{equation}
    \boxed{
    \WMI_L(S)=\CspecL(S)\AspecL(S)
    \frac{\DeffL(S)}{|S|}G_{\mathrm{pair},L}(S)
    =Q_L(S)\frac{\DeffL(S)}{|S|}G_{\mathrm{pair},L}(S).}
    \label{eq:wmi}
\end{equation}
A gain-free organization diagnostic used in state comparisons is
\begin{equation}
    \Oorg(S)=\frac{\DeffL(S)}{|S|}G_{\mathrm{pair},L}(S)
    =\frac{\WMI_L(S)}{Q_L(S)},
    \label{eq:organization-factor}
\end{equation}
with value zero when $Q_L=0$. WMI and $\Oorg$ are summaries of the signature; neither is a universal consciousness index.

Candidate size $k$, horizon $L$, temporal sampling interval, boundary modules, state metric, and any minimum internal shift must be declared before interpreting the score. A practical analysis examines a prespecified size range, uses held-out data and size-matched nulls, and reports sensitivity across the shortest horizons for which the signature and candidate ranking stabilize. The horizon contains $L$ past input samples and $L$ future output samples. Under the indexing in \cref{eq:markov}, a $q$-shifted operator represents input--output separations from $(q+1)\Delta t$ to $(2L+q-1)\Delta t$; $L\Delta t$ alone is not the maximum mediated lag. Raw WMI does not determine a unique core size because useful additions can increase capacity. The dependence of $G_{\mathrm{pair}}$ on the module partition and the nonuniqueness of the WMI scalarization are also unresolved analysis choices. These issues are discussed in \cref{app:parameter-choice}; the conceptual issue of canonical size and boundary is revisited in the Discussion.

\section{Relation to existing measures}
\label{sec:comparison}

The GMW formulation combines features that established measures usually treat separately (\cref{tab:conceptual-comparison}). Graph centrality, rich-club structure, and participation characterize topological brokerage, while communicability sums weighted walks \citep{Freeman1977,GuimeraAmaral2005,VandenHeuvelSporns2011,Katz1953,EstradaHatano2008}. Average and modal controllability quantify one-sided actuation, and sensor/actuator placement optimizes external access to a system \citep{Liu2011,Pasqualetti2014,Summers2016,Manohar2022,Gu2015}. None of these standard formulations requires the selected set to implement a common internal read--transform--write route.

Granger causality and transfer entropy estimate directed predictive or
information-theoretic dependence \citep{Granger1969,Schreiber2000}. In
consciousness research, these measures have been developed into
system-level summaries, most notably causal density, the mean pairwise
conditional Granger causality across a network, which was proposed as a
measure of conscious level that captures the coexistence of dynamical
differentiation and integration \citep{Seth2006,Seth2011}. Causal
density aggregates directed pairwise dependencies, and in that respect
it shares a motivation with the GMW. It does not, however, designate a
mediating subsystem or return the number and alignment of independent
internal mediation modes through a declared boundary, which is the
quantity the boundary Hankel operator is constructed to measure.

Integrated Information Theory 4.0 instead targets intrinsic cause--effect irreducibility, whereas the GMW measures extrinsic mediation across a declared boundary \citep{Albantakis2023}. Information-closure approaches ask whether a scale is predictively autonomous \citep{Bertschinger2008,Chang2020}. These questions are complementary: a subsystem may mediate global access without being maximally irreducible or informationally closed.

\begin{table}[t]
\centering
\caption{Conceptual comparison. A check mark denotes an explicit feature of the usual formulation; a triangle denotes a possible extension rather than the defining object.}
\label{tab:conceptual-comparison}
\scriptsize
\begin{tabularx}{\textwidth}{Y C C C C Y}
\toprule
Method family & Dynamics / walk gain & Bidirectional access & Common internal route & Mode rank / alignment & Primary target \\
\midrule
Degree, rich club, participation & $\times$ & $\triangle$ & $\times$ & $\times$ & Structural hub or connector role \\
Betweenness & shortest paths & $\triangle$ & $\triangle$ & $\times$ & Geodesic brokerage \\
Communicability / Katz & all walks & $\triangle$ & $\triangle$ & $\times$ & Aggregate walk-based access \\
Average/modal controllability & $\checkmark$ & $\times$ & $\times$ & input modes only & One-sided actuation capacity \\
Sensor/actuator placement & $\checkmark$ & $\checkmark$ & usually $\times$ & $\triangle$ & External measurement/control design \\
Granger / transfer entropy & data-dependent & directed pairs & $\triangle$ & usually $\times$ & Predictive or information transfer \\
IIT 4.0 & causal model & intrinsic & intrinsic subset & irreducibility & Intrinsic cause--effect power \\
Information closure / NTIC & stochastic dynamics & boundary-relative & $\triangle$ & information scale & Predictive autonomy \\
GMW boundary mediation & $\checkmark$ & $\checkmark$ & $\checkmark$ & $\checkmark$ & Aligned internal mediation \\
\bottomrule
\end{tabularx}
\end{table}

The synthetic benchmark below places these measures on the same network and shows which confounds each baseline favors.
\section{Synthetic validation and benchmarking}
\label{sec:synthetic}

The synthetic benchmark asks whether the proposed operator recovers a known high-dimensional mediator while rejecting structures that mimic only part of the GMW concept. It also provides the only setting in which the generating dynamics and true candidate are known exactly.

\subsection{Benchmark design}

\begin{figure}[t]
    \centering
    \includegraphics[width=0.78\textwidth]{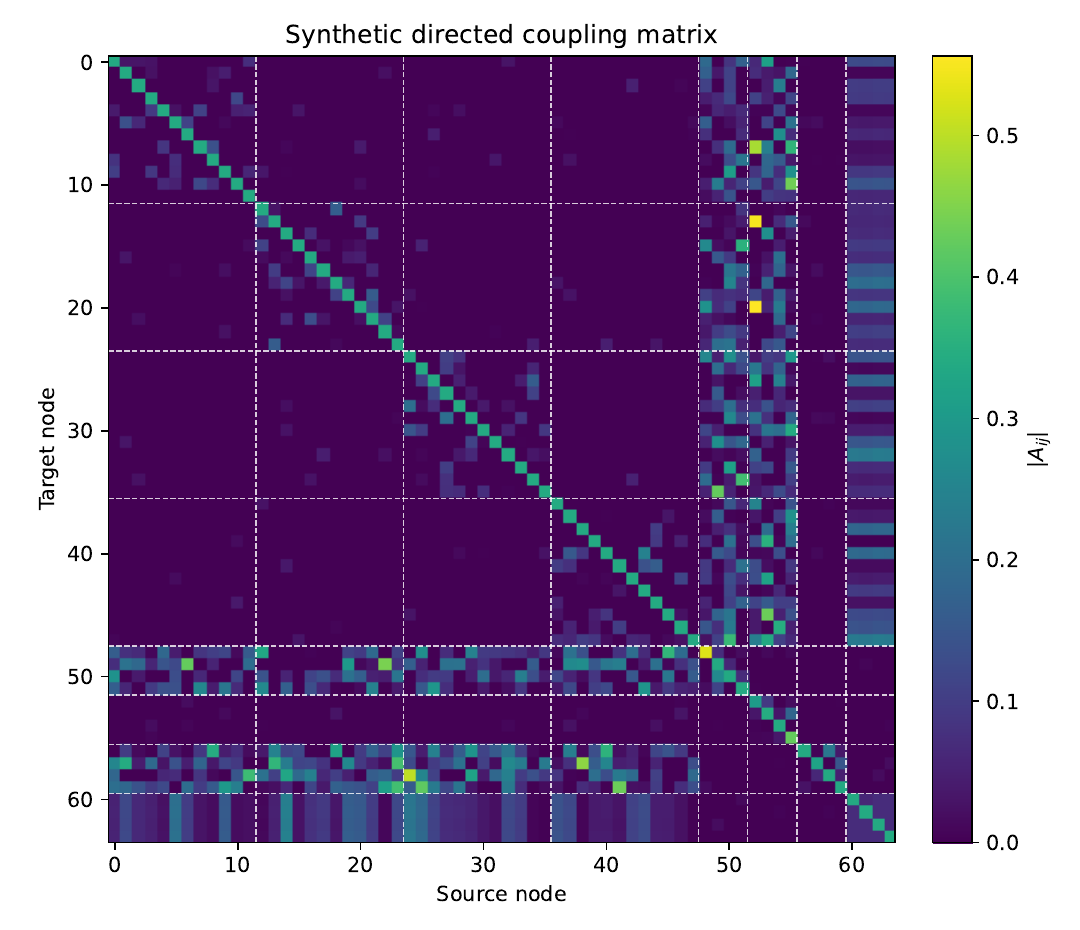}
    \caption{\textbf{Synthetic directed network.} Absolute coupling matrix for the seed-0 benchmark. Four specialist modules occupy nodes 0--47, the planted GMW nodes 48--51, actuator-only nodes 52--55, observer-only nodes 56--59, and the dense low-rank hub nodes 60--63.}
    \label{fig:adjacency}
\end{figure}

The benchmark contained 64 nodes (\cref{fig:adjacency}). Nodes 0--47 formed four specialist modules of 12 nodes with sparse recurrent dynamics and weak intermodule background coupling. Nodes 48--63 instantiated four groups:
\begin{enumerate}[leftmargin=1.8em]
    \item a four-node \textbf{planted GMW} with heterogeneous bidirectional specialist access and internally diverse recurrence;
    \item four \textbf{actuator-only} nodes with strong specialist-directed output and little input;
    \item four \textbf{observer-only} nodes with strong specialist-derived input and little output;
    \item a dense four-node \textbf{degree hub} whose nodes shared nearly collinear input/output profiles.
\end{enumerate}
The split I/O decoy combined two actuator and two observer nodes. A random peripheral control used one node from each specialist module. Weak incidental links made the problem nontrivial and prevented exact block disconnection. The full matrix was rescaled to spectral radius $0.92$; the primary horizon was $L=10$.

\subsection{Competing measures and search strategy}

We computed the following baselines on the same network generated with a fixed random seed (the seed-0 network).
\begin{enumerate}[leftmargin=1.8em]
    \item \textbf{Weighted strength}: total absolute incoming plus outgoing weight.
    \item \textbf{Directed weighted betweenness}: edge length $1/(|A_{ij}|+10^{-6})$.
    \item \textbf{Participation}: distribution of total incident absolute weight across the four specialist modules.
    \item \textbf{Total communicability}: sending plus receiving communicability from $\exp(0.8|A|/\rho)$, where $\rho$ is the spectral radius.
    \item \textbf{Average and modal controllability}: standard structural-network formulas applied to the symmetrized absolute matrix rescaled to spectral radius $0.95$; this convention is reported explicitly because the generative $A$ is directed and signed.
    \item \textbf{Separate placement}: the geometric mean of full-system finite-horizon controllability and observability traces contributed by the selected nodes.
    \item \textbf{External access}: a naive baseline combining controllability of $R$ from $S$ and observability of $R$ at $S$ without requiring internal mediation.
    \item \textbf{WMI}: aligned boundary mediation, differentiated rank, and module-pair breadth.
\end{enumerate}
For additive node metrics, separate placement, and WMI, all $635{,}376$ four-node sets were ranked exactly. External access was retained as a prespecified-candidate comparison because its subset-specific projected Gramian calculation was not part of the optimized exact-ranking suite.

\subsection{Search and robustness analyses}

All $\binom{64}{4}=635{,}376$ four-node sets were evaluated exactly in the seed-0 network. Across ensembles, width-50 beam search was repeated over 50 independently generated networks. Robustness analyses varied the horizon, background coupling, matrix perturbation, and internal routing of the split decoy. Detailed balanced-mode reconstruction and candidate-size analyses are reported in \cref{app:alignment-details,app:parameter-choice}.

\subsection{Successive criteria remove distinct false positives}

\begin{figure}[t]
    \centering
    \includegraphics[width=\textwidth]{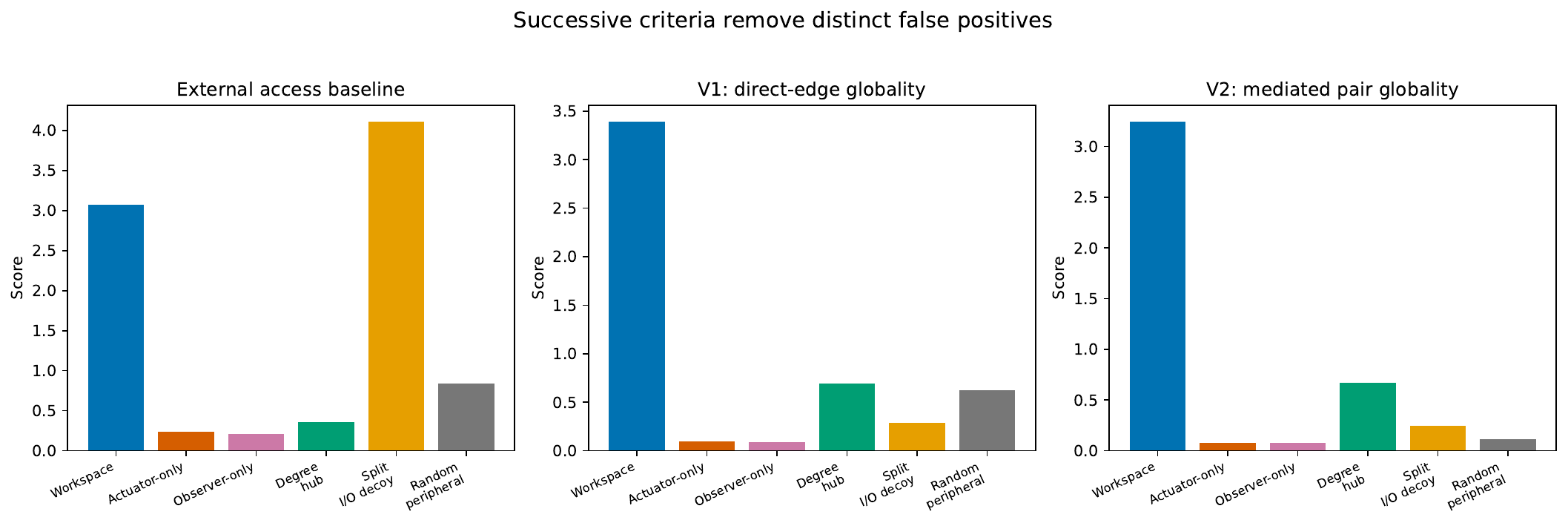}
    \caption{\textbf{Successive criteria and false positives.} External access is defeated by the split input/output decoy. Internal mediation rejects that decoy. Replacing direct-edge module breadth with routed source--target breadth strongly reduces the peripheral set.}
    \label{fig:score-comparison}
\end{figure}

The external-access baseline assigned the split I/O decoy a score of $4.107$, exceeding the planted GMW's $3.074$ (\cref{fig:score-comparison}). The comparison reproduces the conceptual failure: receiver and broadcaster nodes can jointly mimic read/write access without a common internal mechanism.

Internal boundary mediation reduced the split decoy to WMI $0.245$. Replacing direct-edge module breadth with mediated source--target breadth also reduced the random peripheral set to $0.114$, because direct coverage across four modules did not translate into broad routed transformations. The GMW scored $3.242$, compared with $0.078$ for actuator-only, $0.080$ for observer-only, and $0.668$ for the dense hub.

\subsection{Capacity and alignment explain why the decoys fail}

\begin{figure}[t]
    \centering
    \includegraphics[width=\textwidth]{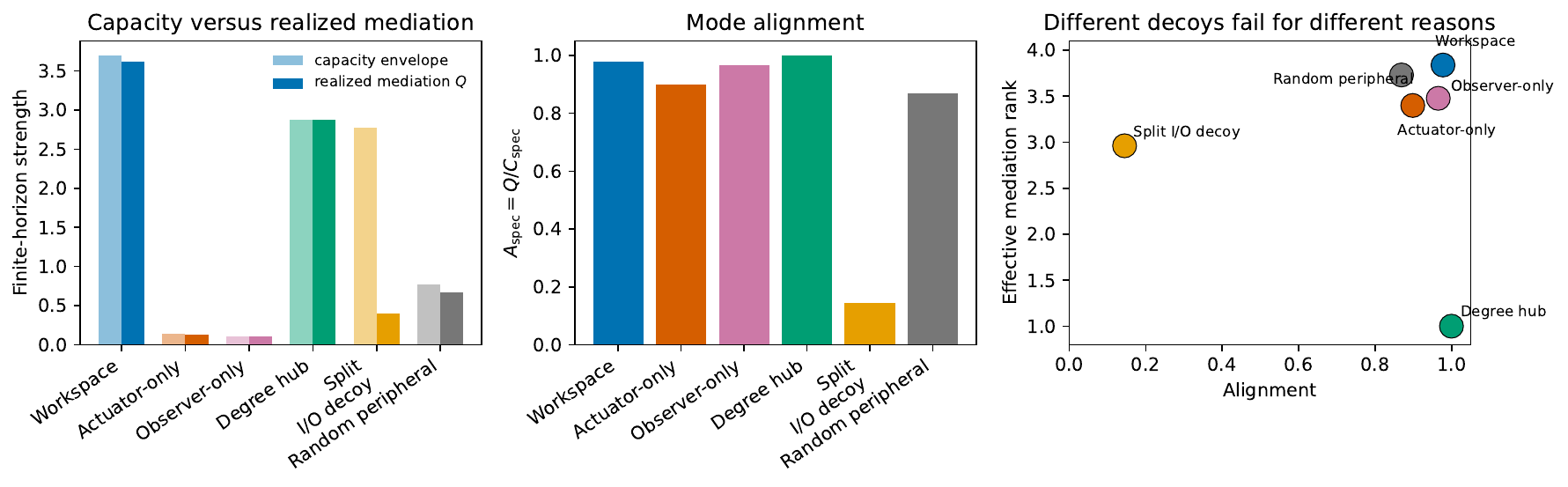}
    \caption{\textbf{Capacity, alignment, and dimensionality.} Left: the split decoy has a large capacity envelope but realizes little of it. Middle: its alignment efficiency is uniquely low. Right: alignment and effective rank separate the split decoy from the aligned but rank-one degree hub; total capacity and module breadth complete the signature.}
    \label{fig:alignment}
\end{figure}

The alignment decomposition made the failure mechanisms explicit (\cref{tab:alignment,fig:alignment}). The split I/O decoy had a capacity envelope $\Cspec=2.775$, comparable to the dense hub, but realized only $Q=0.401$ because $\Aspec=0.145$. Its top-two dominant-subspace principal cosines were $0.0178$ and $0.0009$. Thus, its strong receiving and broadcasting directions were almost orthogonal in the candidate state space.

The degree hub showed the opposite profile. Its capacity and realized strength were nearly identical ($\Cspec=2.882$, $Q=2.881$, $\Aspec=1.000$), but its effective rank was $1.002$. It was a well-aligned but almost one-dimensional mediator. The GMW combined $\Cspec=3.699$, $\Aspec=0.978$, $\Deff=3.839$, and broad module-pair mediation. Actuator-only and observer-only sets had reasonably high alignment ratios but negligible capacity, illustrating why alignment cannot be interpreted without its capacity factor.

\begin{table}[t]
\centering
\caption{Alignment-resolved mediation signature in the seed-0 network. $\Cspec$ is the spectrum-matched capacity envelope and $\Aspec=Q/\Cspec$.}
\label{tab:alignment}
\small
\begin{tabular}{lrrrrrr}
\toprule
Candidate & $\Cspec$ & $\Aspec$ & $Q$ & $\Deff$ & $G_{\mathrm{pair}}$ & WMI \\
\midrule
GMW & 3.699 & 0.978 & 3.616 & 3.839 & 0.934 & \textbf{3.242} \\
Actuator-only & 0.146 & 0.899 & 0.131 & 3.399 & 0.701 & 0.078 \\
Observer-only & 0.112 & 0.965 & 0.108 & 3.477 & 0.860 & 0.080 \\
Degree hub & 2.882 & \textbf{1.000} & 2.881 & 1.002 & 0.926 & 0.668 \\
Split I/O decoy & 2.775 & 0.145 & 0.401 & 2.961 & 0.825 & 0.245 \\
Random peripheral & 0.771 & 0.869 & 0.670 & 3.731 & 0.182 & 0.114 \\
\bottomrule
\end{tabular}
\end{table}

\subsection{Comparison with existing network metrics}

\begin{figure}[t]
    \centering
    \includegraphics[width=\textwidth]{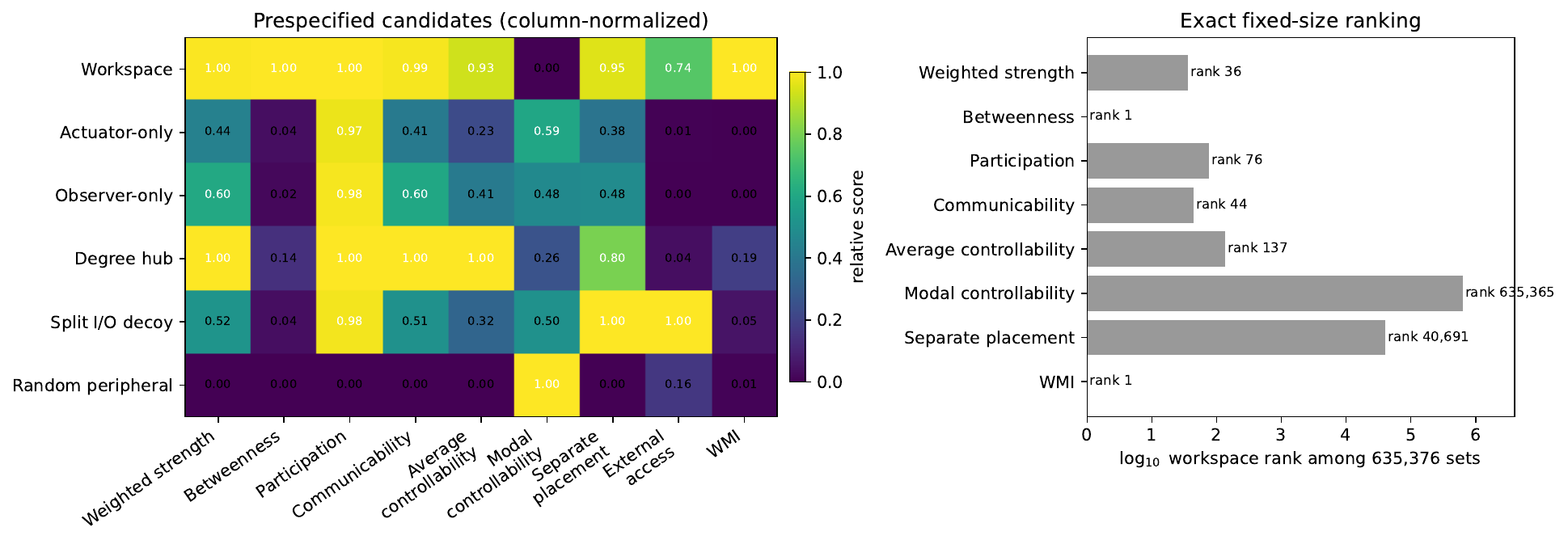}
    \caption{\textbf{Existing-method comparison on the same benchmark.} Left: each column is normalized across the six prespecified candidates, revealing distinct favored decoys. Right: exact rank of the planted GMW among all 635,376 four-node sets. Betweenness also ranks the GMW first here, but does not quantify aligned mediation dimensionality.}
    \label{fig:baselines}
\end{figure}

Standard metrics emphasized different motifs (\cref{fig:baselines,tab:baseline-ranks}). Among the six prespecified candidates, weighted strength, communicability, and average controllability favored the dense hub; modal controllability favored the peripheral set; separate placement and external access favored split input/output sets. Betweenness and participation happened to favor the planted GMW among the named candidates, but this did not generalize to exact subset ranking for participation, and neither metric diagnoses mediation rank or alignment.

Across all four-node sets, the GMW ranked 36th by weighted strength, 76th by participation, 44th by communicability, 137th by average controllability, 635,365th by modal controllability, and 40,691st by separate placement. It ranked first by both betweenness and WMI in this particular network. The two rankings reflect different properties. Betweenness selected the GMW because the synthetic topology routed many shortest paths through it. WMI also showed that the candidate supported several aligned, gain-weighted state-space channels. The best non-GMW, nonoverlapping set under each conventional metric had WMI between $0.071$ and $1.204$, below the planted GMW's $3.242$.

\begin{table}[t]
\centering
\caption{Exact fixed-size comparison. ``Best disjoint'' excludes all planted GMW nodes. External access is shown in the prespecified comparison but was not exhaustively ranked.}
\label{tab:baseline-ranks}
\scriptsize
\begin{tabular}{lrlrr}
\toprule
Metric & GMW rank & Best disjoint set & Its raw score & Its WMI \\
\midrule
Weighted strength & 36 & 60,61,62,63 & 33.218 & 0.668 \\
Betweenness & 1 & 0,18,24,63 & 0.268 & 1.204 \\
Participation & 76 & 52,57,61,63 & 2.986 & 0.555 \\
Communicability & 44 & 60,61,62,63 & 21.176 & 0.668 \\
Average controllability & 137 & 60,61,62,63 & 6.060 & 0.668 \\
Modal controllability & 635,365 & 2,3,4,44 & 3.952 & 0.071 \\
Separate placement & 40,691 & 19,53,55,58 & 38.708 & 0.348 \\
WMI & \textbf{1} & 20,23,24,63 & 1.563 & 1.563 \\
\bottomrule
\end{tabular}
\end{table}

\subsection{Differentiated mediation and module-pair structure}

\begin{figure}[t]
    \centering
    \includegraphics[width=\textwidth]{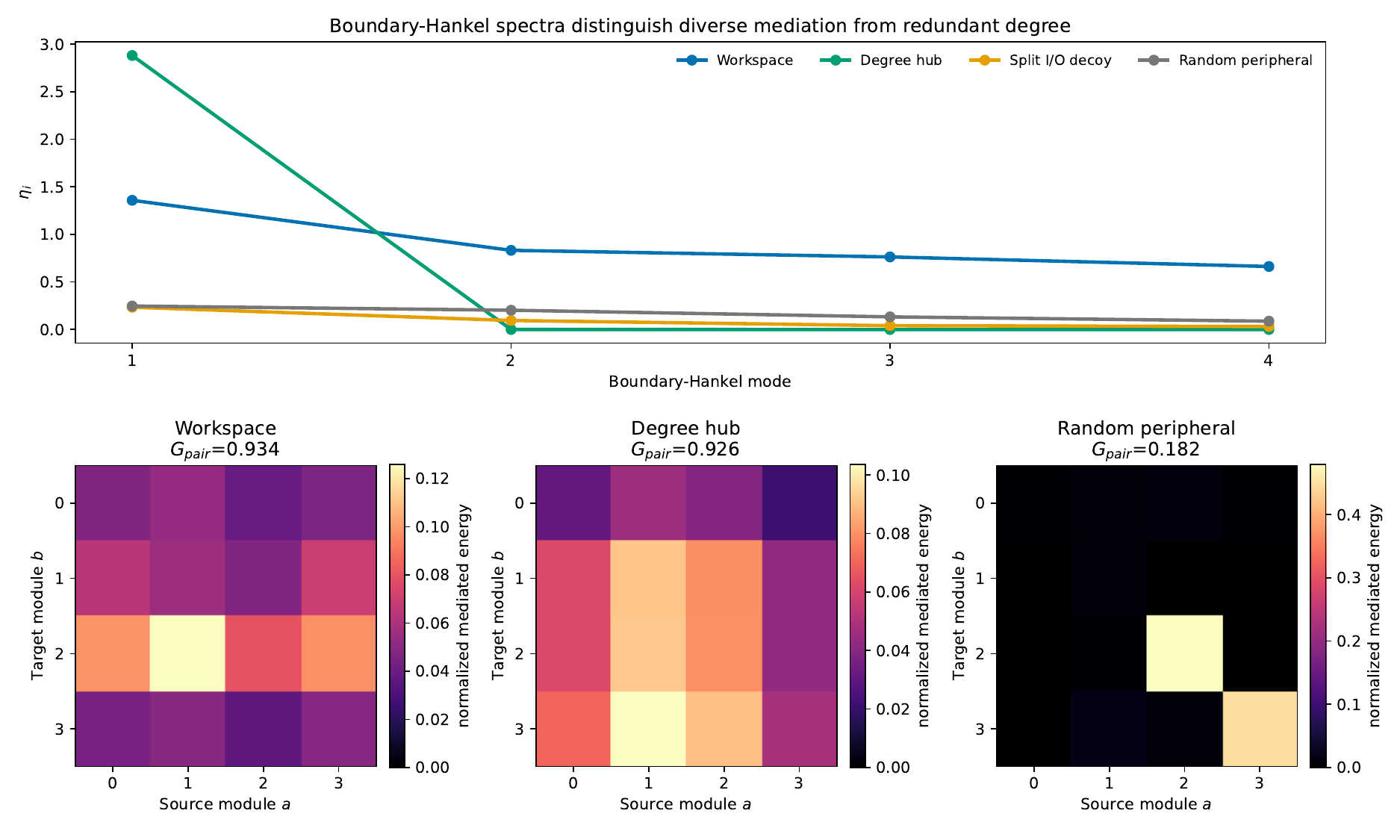}
    \caption{\textbf{Mediation spectrum and module-pair structure.} The dense hub has the strongest first mode but is almost rank one. The GMW contains four substantial modes and distributes mediated energy across source--target specialist pairs.}
    \label{fig:spectrum-pairs}
\end{figure}

The degree hub's spectrum was $(2.881,2.77\times10^{-4},1.08\times10^{-4},6.22\times10^{-5})$, whereas the GMW spectrum contained four substantial modes. Module-pair matrices further separated broad mediation from superficial module coverage: the GMW distributed energy across all 16 source--target pairs; the hub contacted many pairs through one internal mode; the peripheral set concentrated routed energy in a few pairs (\cref{fig:spectrum-pairs}).

\subsection{Exact recovery and robustness}

\begin{figure}[t]
    \centering
    \includegraphics[width=\textwidth]{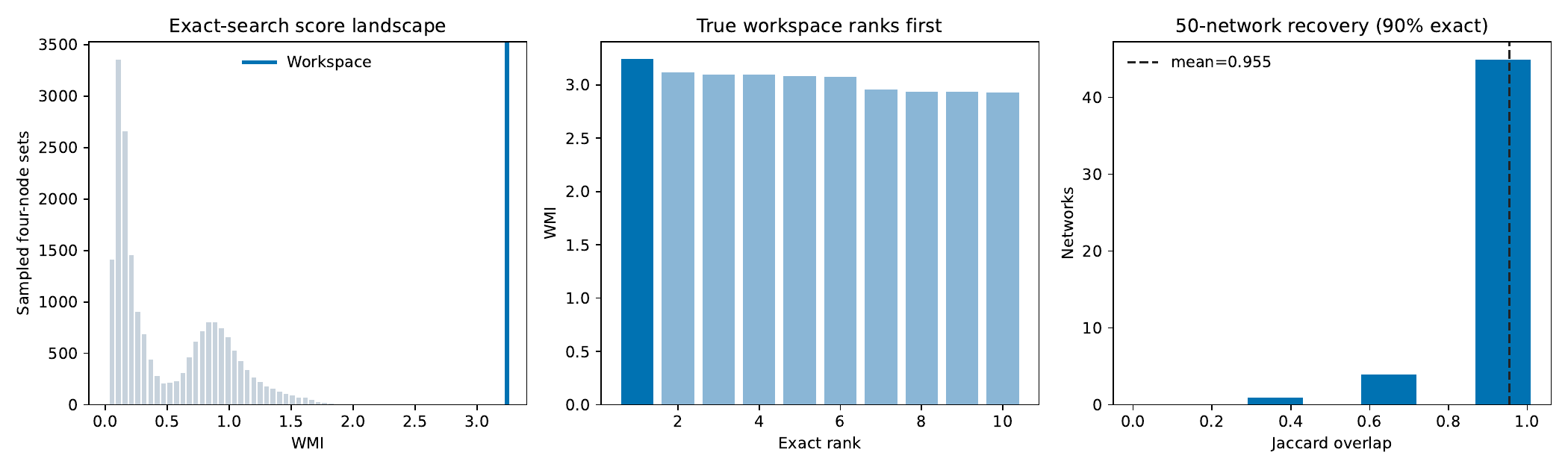}
    \caption{\textbf{Search results.} Exact four-node score landscape, highest-scoring sets, and beam-search overlap across 50 independently generated networks.}
    \label{fig:search}
\end{figure}

Exhaustive evaluation of all 635,376 four-node sets ranked the planted GMW first with WMI $3.24$. The second-ranked set $(48,49,50,63)$ scored $3.12$, and the ten highest sets all contained at least three GMW nodes (\cref{fig:search}). Across 50 independently generated networks, width-50 beam search exactly recovered all four GMW nodes in 45 cases, with mean Jaccard overlap $0.955$.

Recovery was stable across horizons $L=1$--20, background-coupling regimes, and moderate matrix perturbation. Adding the missing observer-to-actuator route to the split decoy increased its WMI continuously. These tests support construct and numerical validity within the synthetic regime, not biological validity.

\begin{figure}[t]
    \centering
    \includegraphics[width=\textwidth]{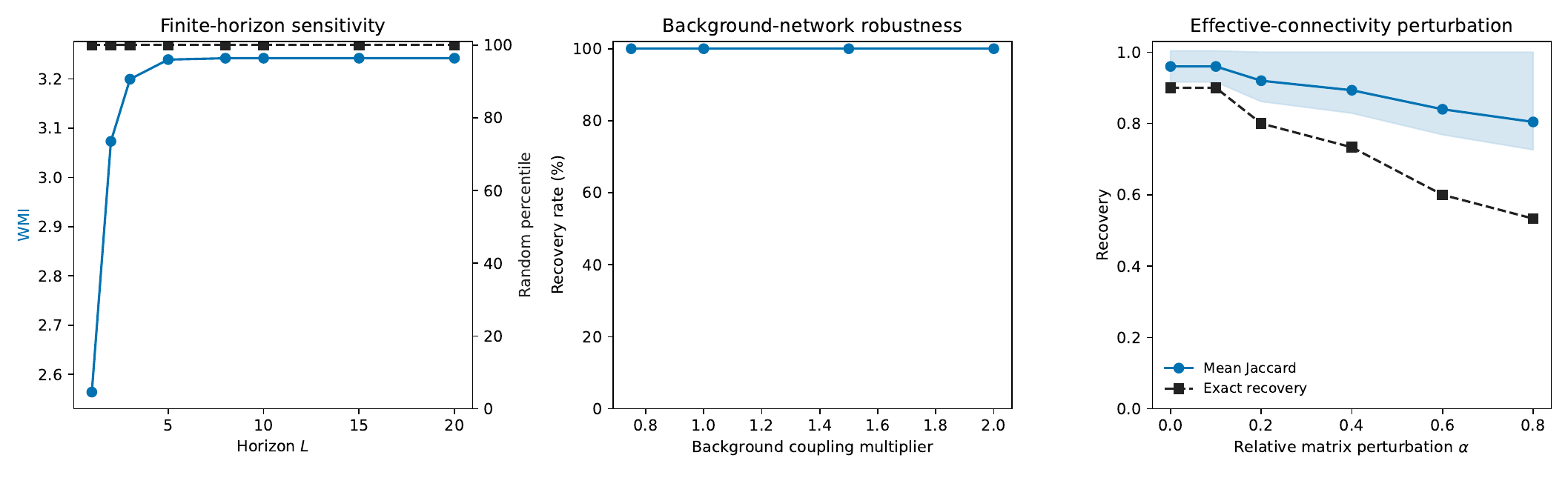}
    \caption{\textbf{Robustness analyses.} Horizon dependence, background-network variation, and paired effective-connectivity perturbations produced gradual changes.}
    \label{fig:robustness}
\end{figure}

\FloatBarrier

The benchmark establishes construct and numerical validity under a known linear generative model. The next section asks how the same input--output principle changes when mediation depends on operating state and perturbation amplitude.

\section{Nonlinear GMW mediation and state-dependent instantiation}
\label{sec:nonlinear-extension}

A nonlinear candidate does not possess one amplitude-independent mediation matrix. Its boundary transfer depends on the reference trajectory, perturbation amplitude, input ensemble, and active regime. The linear operator remains exact for a declared linear realization and becomes the tangent-space building block of the nonlinear formulation. The tangent-space formulation connects the GMW to nonlinear balancing, empirical Gramians, differential balancing, and lifted linear representations \citep{Scherpen1993,Lall2002,KawanoScherpen2017,Proctor2016,BruntonKoopman2022}.

\subsection{Nonlinear boundary map and differential signature}

Partition the nonlinear network into candidate coordinates $z_t$ and remainder coordinates $r_t$,
\begin{align}
    z_{t+1}&=f_S(z_t,r_t),\\
    r_{t+1}&=f_R(r_t,z_t).
\end{align}
Along a reference trajectory $(\bar r_t,\bar z_t)$, define a centered candidate-induced output
\begin{equation}
    h_{S,t}(z)=E_y\left[f_R(\bar r_t,z)-f_R(\bar r_t,\bar z_t)\right],
    \label{eq:nl-counterfactual-output}
\end{equation}
which removes direct $R\rightarrow R$ evolution from the candidate-mediated output. If $\mathcal C^-_{S,L}$ maps a past boundary-input history to the candidate state at the cut and $\mathcal O^+_{S,L,q}$ maps that state to future boundary outputs, the centered nonlinear boundary map is
\begin{equation}
    \mathcal H^{\mathrm{NL}}_{S,L,q}(v)=
    \mathcal O^+_{S,L,q}\!\left[\mathcal C^-_{S,L}(\bar u^-+v)\right]
    -\mathcal O^+_{S,L,q}\!\left[\mathcal C^-_{S,L}(\bar u^-)\right].
    \label{eq:nl-hankel-map}
\end{equation}
The map need not be linear or additive.

Along the same trajectory, let $A_t$, $B_t$, and $C_t$ be the Jacobians of the candidate transition, boundary input, and candidate-induced output. Their finite-horizon variational reachability and observability matrices, $\Reach^-_{S,L}$ and $\Observe^+_{S,L,q}$, give
\begin{equation}
    D\mathcal H^{\mathrm{NL}}_{S,L,q}(0)
    =\Observe^+_{S,L,q}\Reach^-_{S,L}.
    \label{eq:nl-differential-factorization}
\end{equation}
The singular values
\begin{equation}
    \eta_i^{\mathrm{diff}}(S\mid\bar\tau,L,q)
    =\sigma_i\!\left(D\mathcal H^{\mathrm{NL}}_{S,L,q}(0)\right)
    \label{eq:nl-local-spectrum}
\end{equation}
form a trajectory-conditioned differential mediation spectrum. The linear GMW signature can be evaluated locally along a trajectory. A smooth internal coordinate change leaves the total differential boundary operator unchanged when the corresponding metric is transformed consistently; the derivation and coordinate statement are given in \cref{app:nonlinear-details}.

\subsection{Finite amplitude and the instantiating coalition}

A local derivative can miss a gate that opens only after a finite perturbation. For perturbation direction $v\sim\mu$, the best linear secant operator at amplitude $\epsilon$ is
\begin{equation}
    K_{\epsilon,\mu}
    =\mathbb E_\mu\!\left[\frac{\mathcal H^{\mathrm{NL}}(\epsilon v)}{\epsilon}v^\top\right]\Sigma_v^\dagger.
    \label{eq:nl-secant}
\end{equation}
Its spectrum is explicitly conditioned on amplitude and input ensemble. Higher derivatives can quantify cross-source interactions, while nonsmooth systems require an active-set or finite-amplitude convention.

For a fixed candidate size and matched family $\mathcal C_k$, define the coalition currently instantiating the GMW by
\begin{equation}
    \widehat S_t=\arg\max_{S\in\mathcal C_k}
    \WMI^{\mathrm{diff}}_{L,q}(S\mid\bar\tau_t).
    \label{eq:nl-instantiating-coalition}
\end{equation}
The instantiating coalition can change with context even when the anatomical network is fixed. Candidate size and boundary selection remain independent modeling choices.

\subsection{Synthetic nonlinear validation}

\begin{figure}[t]
    \centering
    \includegraphics[width=\textwidth]{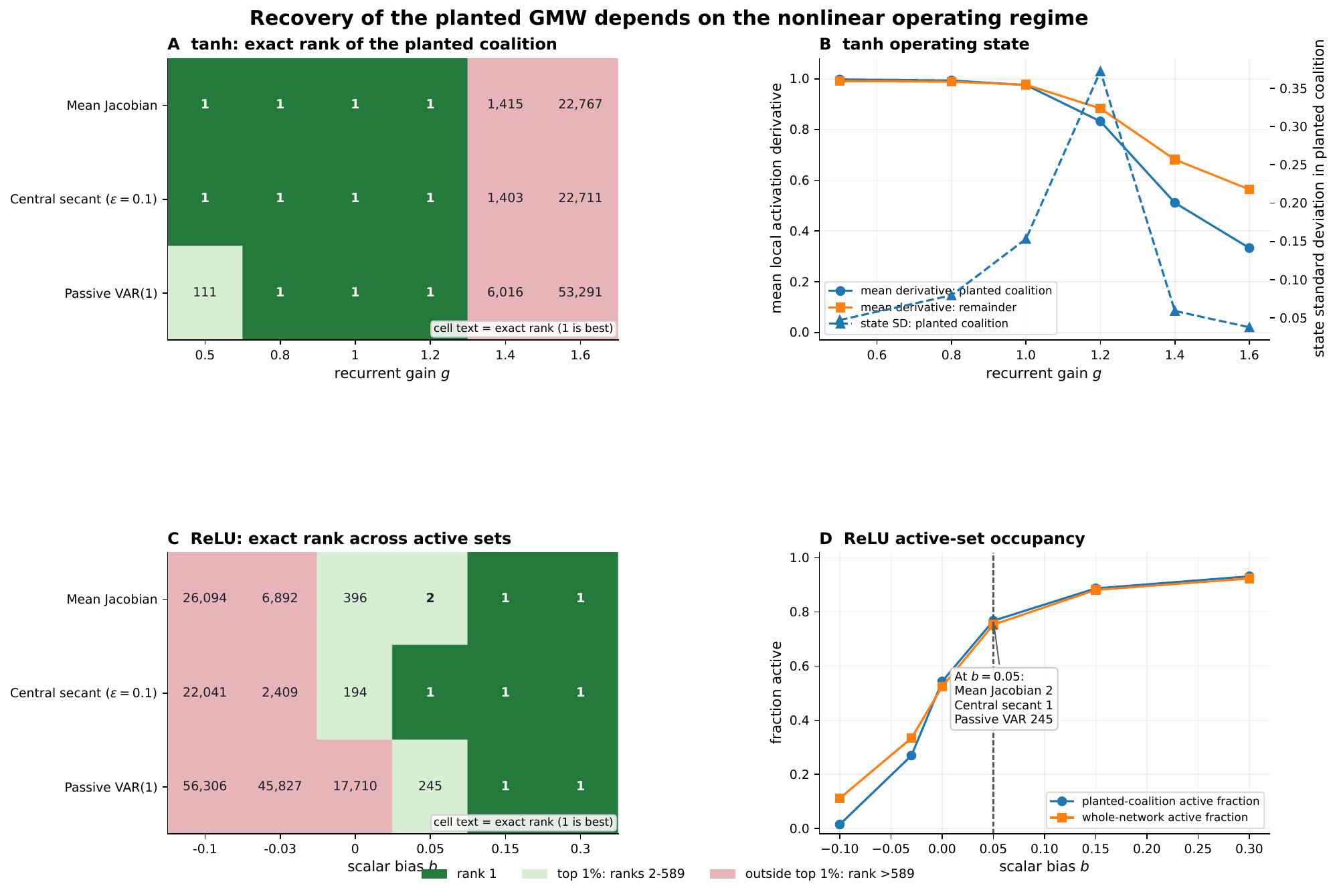}
    \caption{\textbf{Nonlinear operating regimes and planted-candidate recovery.} (A,C) Exact rank of the planted candidate among $58{,}905$ four-node candidates for tanh gain and ReLU bias. Colors distinguish rank $1$, the remaining top $1\%$ (ranks $2$--$589$), and ranks below the top $1\%$. (B) Mean activation derivatives in the planted candidate and the remainder, together with candidate-state standard deviation. (D) Active fractions in the planted candidate and the whole network; the dashed line marks $b=0.05$.}
    \label{fig:activation-benchmark}
\end{figure}

Standard activation functions were first applied to a network with a planted mediator. Moderate tanh dynamics preserved the planted coalition, whereas high gain changed the active operator and eventually favored another candidate (\cref{fig:activation-benchmark}A--B). A separate sample-size control at $g=0.5$ showed that passive VAR rank improved from $36{,}319$ to $1$ when the number of observed transitions increased from $500$ to $32{,}000$; this was a finite-sample estimation effect, not a manipulation of the noise ratio. The mean Jacobian and central secant use access to model derivatives or controlled perturbations and do not incur the same passive-estimation burden. At high gain, even these operators ceased to recover the planted coalition, indicating that saturation had changed the functional operator itself.

ReLU produced a different operating-state dependence (\cref{fig:activation-benchmark}C--D). Near the active-set kink at $b=0.05$, the central secant ranked the planted candidate first, the mean Jacobian second, and the passive VAR $245$th. The active fractions in panel D locate this transition, while panels A and C report candidate recovery. For hard thresholds, a pathwise Jacobian was uninformative but a noise-averaged ensemble operator recovered the coalition in a switching regime; the nonsmooth benchmark is reported in \cref{fig:nl-activation-extra}.

\begin{figure}[t]
    \centering
    \includegraphics[width=0.98\textwidth]{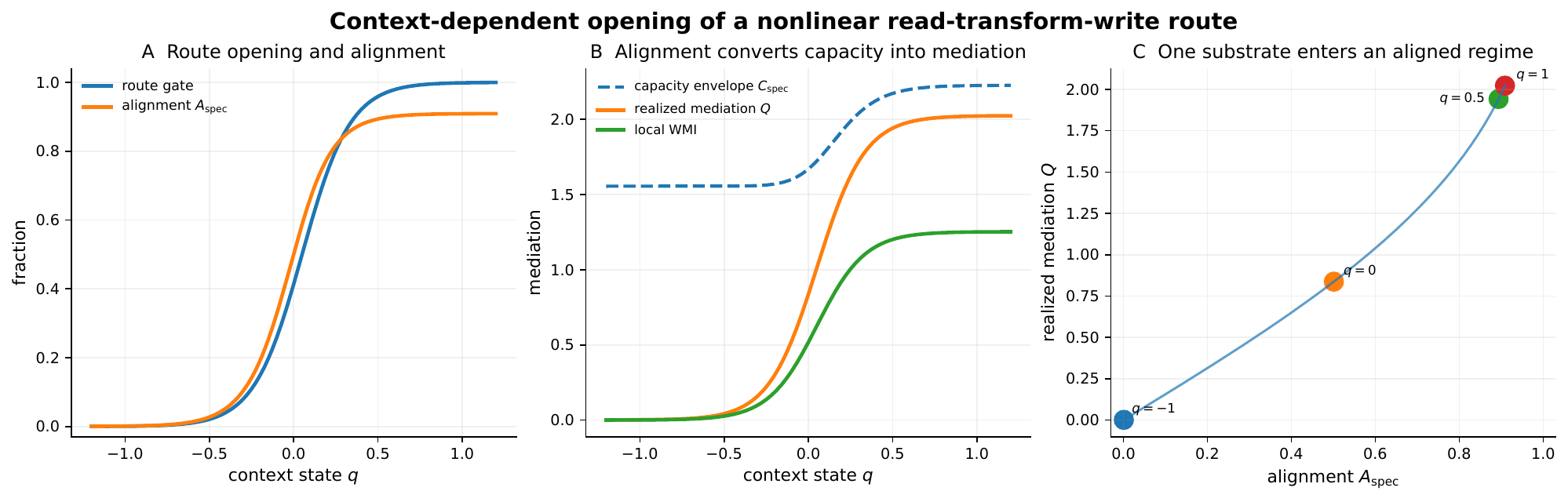}
    \caption{\textbf{Context-dependent opening of a nonlinear mediation route.} (A) A latent context variable opens the internal route and increases input--output alignment. (B) Potential capacity $C_{\mathrm{spec}}$ is already nonzero in the closed regime, while realized mediation $Q$ and local WMI remain near zero until the reachable and observable directions align. (C) The same anatomical substrate moves through the $A_{\mathrm{spec}}$--$Q$ plane as context changes; marker size is proportional to effective dimensionality. The example separates anatomical availability from state-dependent functional recruitment.}
    \label{fig:nl-route-opening}
\end{figure}

\begin{figure}[t]
    \centering
    \includegraphics[width=0.98\textwidth]{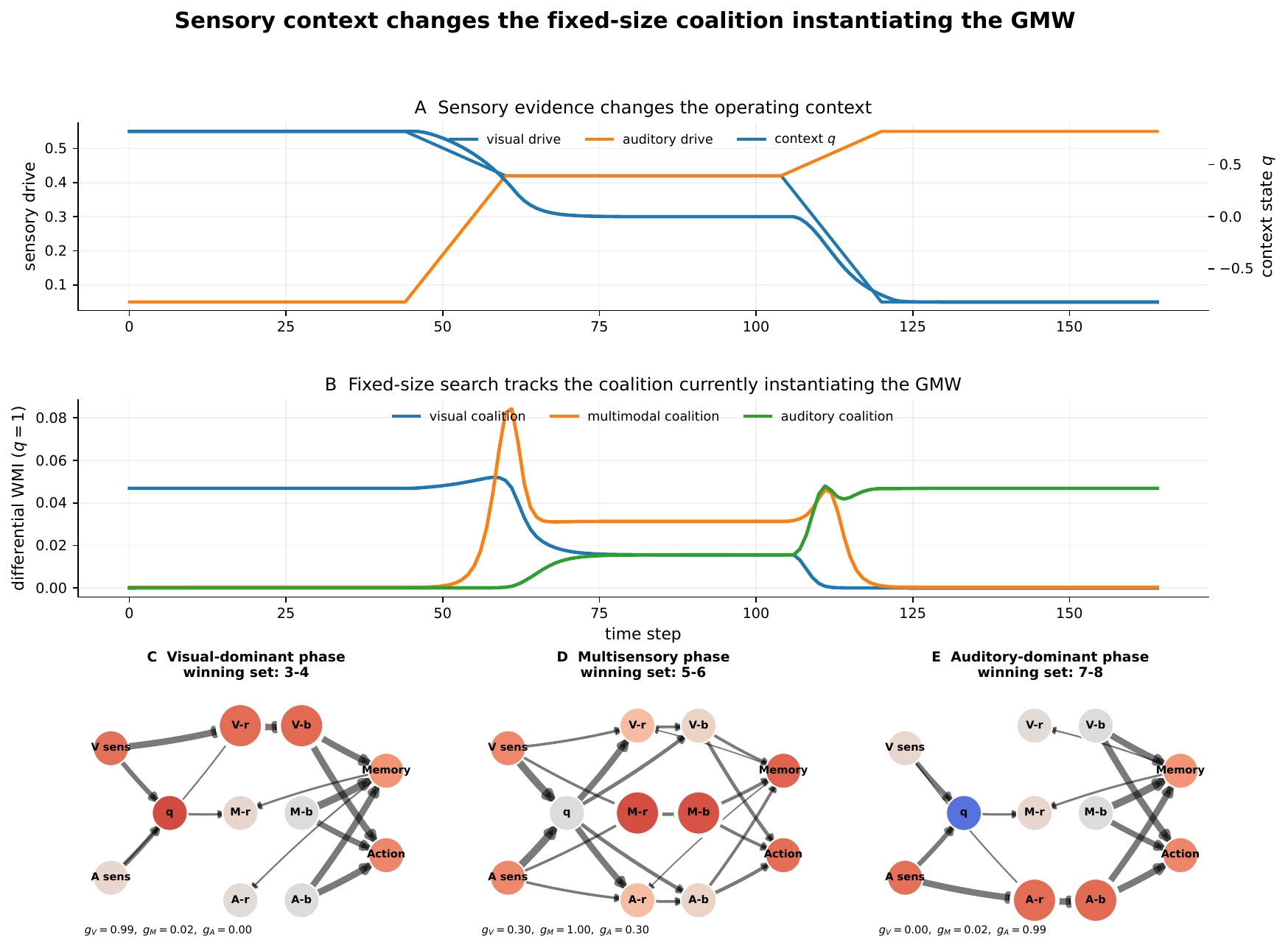}
    \caption{\textbf{Sensory context changes the coalition currently instantiating the GMW.} (A) Visual and auditory drives alter a latent context state. (B) Differential WMI is evaluated for three same-size candidate coalitions; the maximizing coalition changes from visual to multimodal to auditory as the operating context changes. (C--E) Snapshots show the corresponding active Jacobian and winning node pair in visual-dominant, multisensory, and auditory-dominant phases. The anatomical network remains fixed, so the changing winner reflects state-dependent recruitment and not structural relocation.}
    \label{fig:nl-coalition-switching}
\end{figure}

A context-gated system separated receiver and broadcaster states while a latent variable opened their internal route. Potential capacity was already present when the gate was closed, but realized mediation remained small until the receive and send modes aligned (\cref{fig:nl-route-opening}). A second simulation switched the maximizing fixed-size coalition between visual, multimodal, and auditory candidates as sensory context changed (\cref{fig:nl-coalition-switching}). Together, these examples distinguish three questions: whether the anatomical route exists, whether the current operating regime makes it functional, and whether available data identify the active operator.

Detailed finite-amplitude, higher-order, nonsmooth, and benchmark specifications are reported in \cref{app:nonlinear-details}.
\FloatBarrier

\section{Macaque ECoG: GMW organization across wakefulness, deep anesthesia, and recovery}
\label{sec:macaque-empirical}

The macaque analysis in this section is a preliminary application rather than a confirmatory test of global workspace theory. Its purpose is to assess whether the GMW quantities can be estimated from ECoG recordings, how the components change across a well-characterized state transition, and which estimation problems become visible in practice. Ketamine--medetomidine anesthesia provides a useful test because cortical activity can become slow, coherent, and highly predictable. Interaction gain and mode organization can therefore change in different directions.

\subsection{Data and selection-safe analysis}

We analyzed 11 experiment days from four macaques in the NeuroTycho anesthesia resource and associated large-scale information-flow studies \citep{Yanagawa2013,Tajima2015,NeuroTychoTask75}. Every day contained awake and deep-anesthesia recordings. Nine days from three monkeys also contributed the confirmed full-state recovery analysis. The biological sample size is four animals; days, temporal blocks, folds, and candidate sizes are repeated measurements within animals.

The 128-contact recordings were converted to animal-specific, no-contact-reuse bipolar representations. Two animals yielded 63 independent variables, and the completed geometry-aware rematching of the other two yielded 64 variables, using all 128 contacts exactly once. All incidence matrices were full row rank. State-specific ridge dynamics were fitted at a 25-ms lag with awake-derived scaling held fixed across states. Candidate sizes $k=3,4,5$ were prespecified. WMI was used only to rank candidates within size, and state comparisons used the full GMW signature. Within-day awake cross-fitting and same-animal leave-one-day-out (LODO) transfer separated candidate discovery from state evaluation. The principal horizon was $L=4$ and the internal shift was $q=1$, representing input--output separations from 50 to 200~ms. Detailed preprocessing, selection procedures, montage construction, and sensitivity analyses are reported in Appendices~\ref{app:macaque-methods} and~\ref{app:macaque-robustness}.

\subsection{Deep anesthesia increases gain while reducing alignment}

Short-lag prediction improved on every experiment day: median held-out $R^2$ increased from $0.253$ in awake data to $0.890$ during deep anesthesia. Same-animal LODO analysis also showed robust increases in realized mediation and potential capacity. At $k=4$, the animal-balanced deep/awake ratio was $2.13$ [1.46, 3.57] for $Q$ and $2.36$ [1.58, 4.25] for $\Cspec$, while $\Aspec$ decreased to $0.91$ [0.84, 0.98]. At this size, $\Deff/k$, $G_{\mathrm{pair}}$, and $\Oorg$ had point estimates of $0.95$, $0.98$, and $0.93$, with intervals that included one. Reductions in differentiated organization were clearer at $k=5$: $\Aspec=0.89$ [0.80, 0.99], $\Deff/k=0.91$ [0.81, 0.99], $G_{\mathrm{pair}}=0.83$ [0.69, 1.06], and $\Oorg=0.75$ [0.56, 1.06], while top-mode share increased to $1.21$ [1.00, 1.46]. Across $k=3$--5, $Q$ and $\Cspec$ increased in all four animals and $\Aspec$ was below one at every size. The most stable state effect was therefore an increase in gain and capacity accompanied by lower input--output alignment. Lower dimensionality, routed breadth, and gain-free organization were size dependent and most apparent for the largest candidate sets.

Raw WMI remained above one at all sizes ($2.29$, $1.98$, and $1.77$ for $k=3,4,5$) because the gain increase outweighed the reductions in other components. Strong and predictable dynamics can therefore produce a high scalar WMI without preserving the same degree of mode alignment or differentiated routing. Full size-specific estimates and held-out-selection analyses are reported in Appendix~\ref{app:macaque-methods}.

\FloatBarrier
\subsection{Cortical distribution of repeatedly selected candidate sites}

\begin{figure}[H]
    \centering
    \includegraphics[height=0.50\textheight,keepaspectratio]{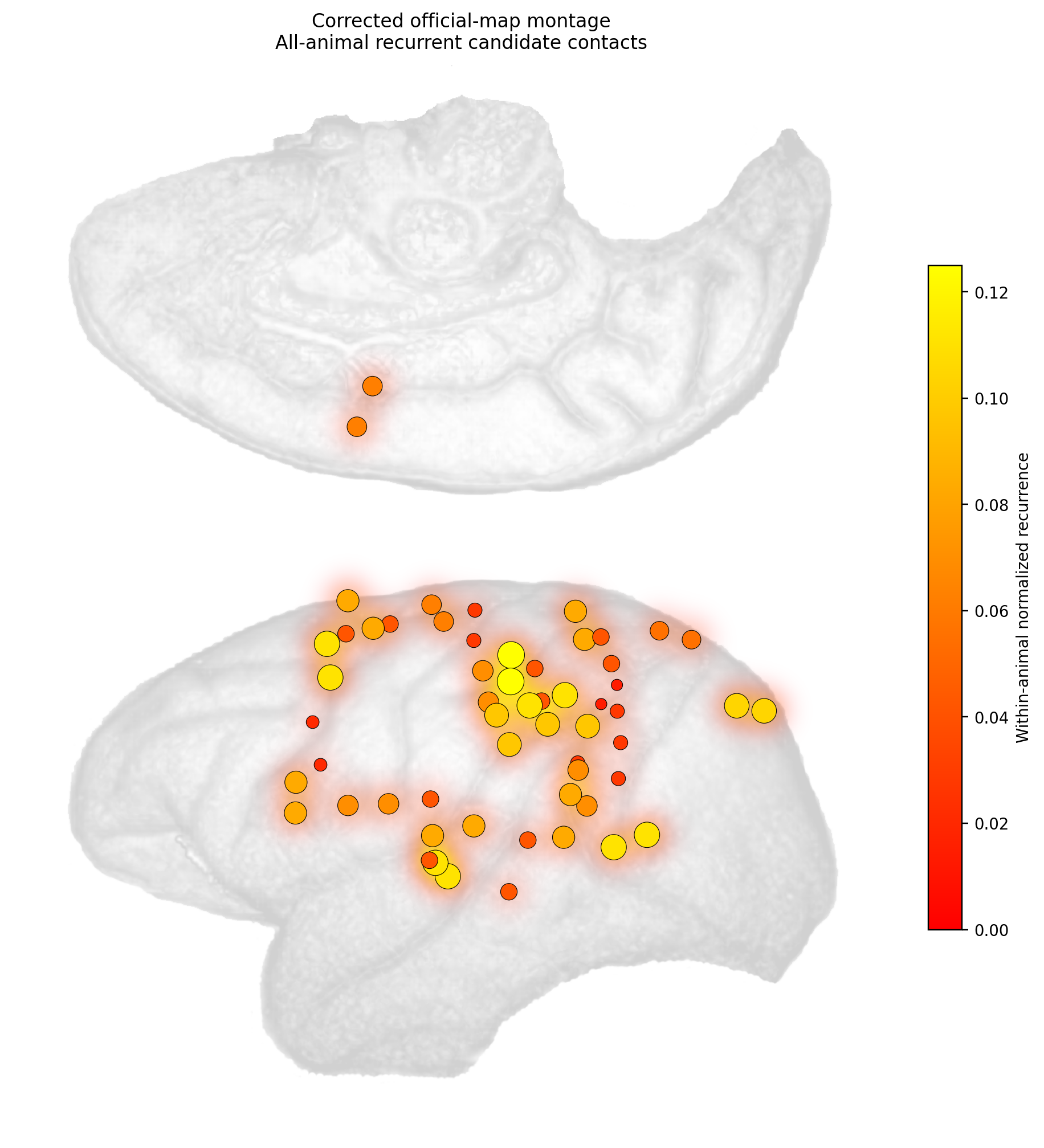}
    \caption{\textbf{Broad cortical distribution of repeatedly selected awake-candidate sites.} Candidate recurrence was normalized within animal, averaged across $k=3$--5, and projected panel-wise onto a common two-dimensional display. Warmer and larger markers indicate higher within-animal normalized recurrence. The display is intended for spatial orientation and is not a stereotactic registration.}
    \label{fig:macaque-localization}
\end{figure}

Because exact candidate composition varied across days, we examined whether recurrently selected sites nevertheless occupied a broad cortical scaffold. Selection frequency was computed within animal across same-animal LODO folds, averaged across $k=3$--5, and displayed on a common two-dimensional electrode-layout template for spatial orientation (\cref{fig:macaque-localization}). Frequently selected sites were distributed across frontal and premotor, parietal and sensorimotor, temporal, and other association-related sectors. No single focal region dominated the cross-animal summary.

The pooled display is normalized within animal and is not a stereotactic registration. Subject-specific layouts are provided in Appendix~\ref{app:macaque-robustness}. The spatial result supports a distributed cortical scaffold but does not identify a unique anatomical GMW.

\subsection{Descriptive state trajectories and component-specific recovery}

\begin{figure}[H]
    \centering
    \includegraphics[width=0.99\textwidth]{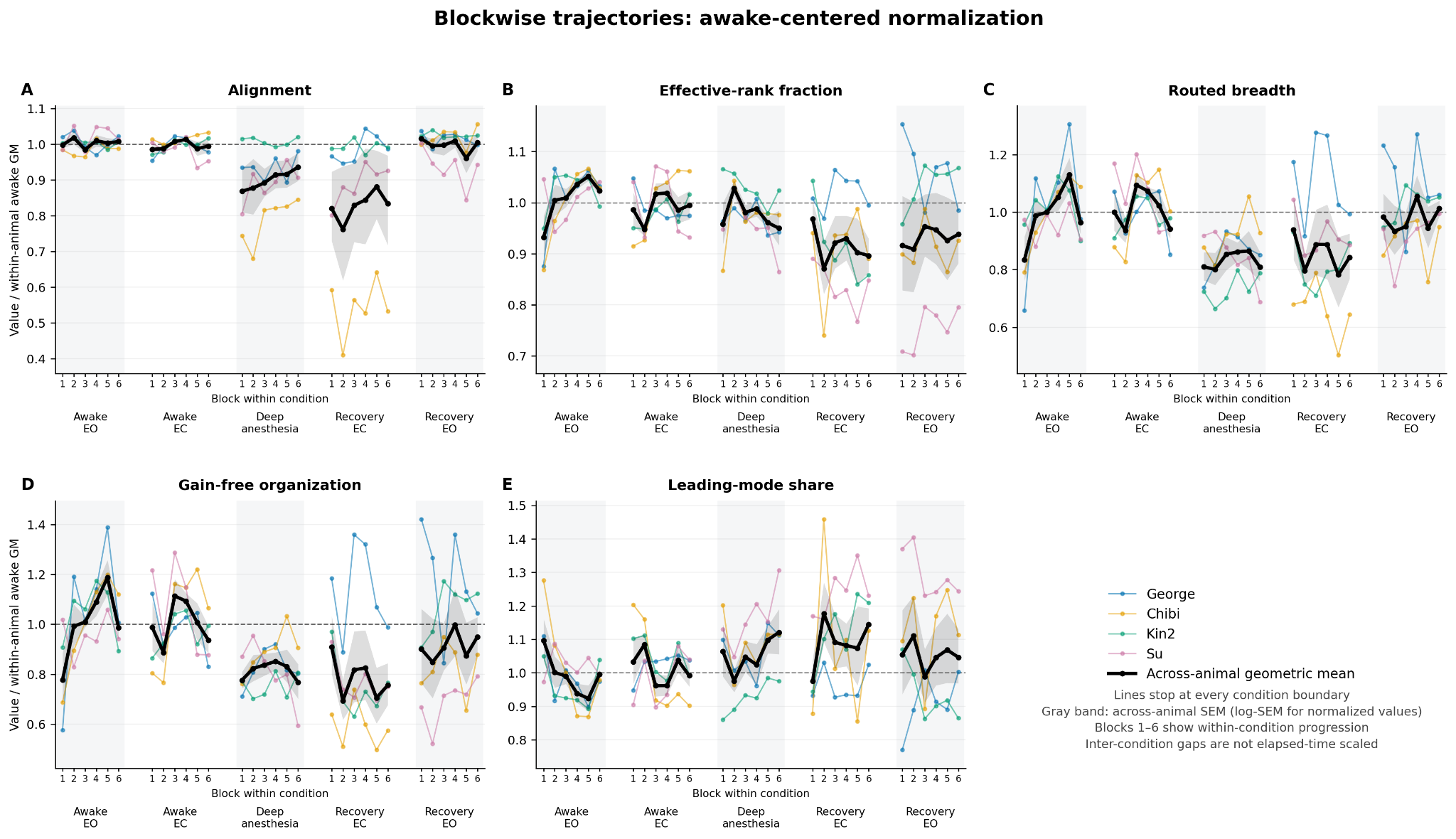}
    \caption{\textbf{Corrected-montage blockwise trajectories with awake-centered normalization.} Six equal-duration blocks are shown for each condition. Colored lines show animal-level trajectories, the black line is the equal-animal geometric mean, and the gray band is the log-scale standard error across animal means. Each animal is normalized so that the combined awake eyes-open and eyes-closed geometric mean equals one. Lines stop at condition boundaries.}
    \label{fig:macaque-all-animal-trajectories}
\end{figure}

We constructed a descriptive blockwise summary using the same-animal LODO candidates. Each state was divided into six equal-duration blocks. For every animal, day, candidate size, and metric, values were first divided by the geometric mean across the 12 awake blocks (eyes open and eyes closed). After averaging across days and $k=3$--5 within animal, each animal was rescaled so that the combined awake eyes-open and eyes-closed geometric mean equaled one. The black trajectory in \cref{fig:macaque-all-animal-trajectories} is the equal-animal geometric mean. State gaps are retained because the blocks do not form a continuous pharmacokinetic time series. The plot is descriptive and is not used as a formal population-level test.

In this all-available-block summary, the deep/awake animal-balanced ratios were $0.90$ for $\Aspec$, $0.98$ for $\Deff/k$, $0.83$ for $G_{\mathrm{pair}}$, and $0.81$ for $\Oorg$, while leading-mode share increased to $1.05$. Recovery trajectories were more heterogeneous. Alignment returned close to its awake value during eyes-open recovery, while effective-rank fraction and gain-free organization remained below the awake average in the blockwise display. Full-state recovery estimates, restricted to nine confirmed recovery days from three animals, are reported in Appendix~\ref{app:macaque-robustness}. The corresponding unnormalized metric trajectories are also shown there.

\FloatBarrier

\section{Discussion}
\label{sec:discussion}

\subsection{From global-workspace functions to the GMW signature}

The Global Mediation Workspace assigns separate dynamical quantities to several functional claims made by global workspace theory. In the cognitive formulation, specialist systems must be able to place information into a limited-capacity workspace; workspace states must then become available to perceptual, mnemonic, evaluative, and action systems; and the state that is broadcast should preserve what was admitted closely enough to support coordinated use \citep{Baars1988,Dehaene1998,DehaeneNaccache2001,DehaeneChangeux2011,Mashour2020}. The control-theoretic quantities introduced here separate these requirements without assuming that one scalar captures all of them (\cref{tab:gwt-gmw-map}).

\begin{table}[htbp]
\centering
\caption{Functional correspondence between global workspace theory and the GMW formulation. The mathematical quantities operationalize aspects of the cognitive requirements; they are not asserted to exhaust those requirements.}
\label{tab:gwt-gmw-map}
\small
\begin{tabularx}{\textwidth}{Y Y Y}
\toprule
Global-workspace requirement & GMW object & Operational interpretation \\
\midrule
Specialists can write to the workspace & $\Wc$, reachability spectrum $\Lambda_c$ & Boundary activity can drive distinct internal directions of the candidate \\
Workspace content can be broadcast to specialists & $\Wo$, observability spectrum $\Lambda_o$ & Candidate-state differences produce distinguishable effects in the remainder \\
What is admitted is what becomes available & $\Aspec$ & Reachable and observable directions are aligned through the same internal modes \\
The workspace has usable two-sided capacity & $\Cspec$ and $Q=\Cspec\Aspec$ & Potential receive/send capacity is converted into realized mediation \\
Arbitrary specialist pairs can be linked through a common blackboard & $G_{\mathrm{pair}}$ & Mediation is distributed across source--target module pairs, not only across incident edges \\
The workspace is limited but not undifferentiated & $k\ll n$ and $\Deff/k$ & A compact bottleneck retains several distinguishable mediation modes \\
Global availability unfolds over a finite time & $L$, $q$, and $\Delta t$ & The operator uses $L$ past and $L$ future samples, with an explicit range of input--output lags \\
\bottomrule
\end{tabularx}
\end{table}

The alignment term gives a dynamical interpretation to the representational requirement of compatibility. The Global Latent Workspace proposes that heterogeneous specialist representations become mutually usable through translation into a shared latent format \citep{VanRullenKanai2021}. The two proposals are not identical: one concerns representational translation, while the other concerns input--output dynamics. Their common requirement is compatibility. In the GMW, $\Aspec$ is high when the internal directions reached from specialist activity are also the directions that can be expressed back to specialist systems. Large receive and send capacities with low $\Aspec$ describe a system that can absorb one set of variations and broadcast another, which fails to preserve the functional continuity implied by a shared workspace.

Limited capacity and differentiated mediation are complementary, not contradictory. The candidate size $k$ specifies the compression bottleneck relative to the full system, whereas $\Deff$ measures how many distinguishable routes are supported inside that bottleneck. A GMW may be low dimensional relative to the whole network while remaining high dimensional relative to its own size. In the animal-balanced ECoG trajectories, $k$ was fixed while deep anesthesia reduced $\Deff/k$ and increased leading-mode share. The bottleneck remained present, but its differentiated repertoire became more concentrated. The distinction between $k$ and $\Deff$ separates functional compression from collapse toward a dominant mode.

\subsection{A recurrent boundary and an operator-level account of ignition}

The same remainder $R$ appears on both sides of the GMW map. Specialist systems are not divided permanently into an audience and a set of performers; each can supply activity to the candidate, receive mediated effects, or occupy both roles at different moments. The operation
\begin{equation}
    R_a\longrightarrow S\longrightarrow R_b
\end{equation}
can connect specialist pairs that need not possess a direct effective route. This recurrent boundary distinguishes the GMW from a feedforward relay and from separate sensor-placement and actuator-placement problems. Many-to-many routed breadth is measured separately from degree.

The nonlinear extension suggests an operator-level hypothesis for ignition. Suppose that a candidate already has substantial potential capacity $\Cspec$, but the directions reached by incoming activity are poorly aligned with the directions that can be expressed outward. Realized mediation is then small because
\begin{equation}
    Q=\Cspec\Aspec.
\end{equation}
An ignition-like event can be described as a rapid state- or amplitude-dependent rise in $\Aspec$, often accompanied by increases in $\Deff$ and $G_{\mathrm{pair}}$, that converts latent two-sided capacity into differentiated, system-wide mediation. The context-gated simulation in \cref{fig:nl-route-opening} isolates this mechanism: a substantial two-sided capacity envelope exists before the gate opens, while realized mediation rises only when incoming and outgoing directions become aligned. The finite-amplitude analysis in Appendix~\ref{app:nonlinear-details} supplies the complementary perturbational signature (\cref{fig:nl-finite-appendix}): inputs below the gate-opening range fail to recruit the route, intermediate amplitudes produce a steep increase in mediation, and larger amplitudes can saturate. Together, these examples reproduce two defining features of workspace ignition---threshold dependence and abrupt conversion of local processing into broadly consequential processing---without defining ignition by raw activity amplitude, a frequency band, or a fixed anatomical locus. In this account, ignition is a change in the boundary operator through which available capacity becomes usable global access; it is not defined by a late response, a specific frequency band, or a fixed anatomical site.

The nonlinear formulation also leads to predictions for classic access paradigms. Under backward masking, a subthreshold perturbation may activate local processing without opening the finite-amplitude route needed for broad mediation; increasing stimulus strength or target--mask separation should produce a steep rise in the differential or secant GMW signature \citep{DelCul2007}. During the attentional blink, a coalition already mediating the first target should transiently reduce the incremental aligned capacity or differentiated breadth available to the second target, even when early sensory processing of that target survives \citep{SergentDehaene2004}. Conscious access should be accompanied by a transient increase in $\Aspec$, $\Deff$, and routed breadth during the late distributed response classically associated with ignition and long-range exchange \citep{Gaillard2009,Mashour2020}. These predictions concern the operator estimated around an event; they do not equate the GMW with the P3b or with any single electrophysiological component.

\subsection{Access consciousness, phenomenal consciousness, and asymmetric mediation}

Block distinguished phenomenal consciousness---what it is like to undergo an experience---from access consciousness, the availability of information for reasoning and the rational guidance of speech and action \citep{Block1995,Block2007,FazekasOvergaard2018}. Whether phenomenal content can exist without cognitive access has remained one of the central disputes in consciousness research. Overflow and recurrent-processing accounts appeal to partial-report performance, high-capacity visual representations, unattended summary information, and local recurrent processing to argue that experience may exceed what is available for report or working memory \citep{Landman2003,Block2011,Lamme2006,Bronfman2014,Amir2023}. Competing interpretations invoke partial or hierarchical access, unconscious representations, richer notions of cognitive availability, or the methodological difficulty of establishing an experience that is inaccessible by hypothesis \citep{Kouider2010,CohenDennett2011,Naccache2018,Overgaard2018,Phillips2018}. No single behavioral paradigm currently resolves this disagreement.

The GMW does not decide which side is correct, but it describes one possible dynamical form of the proposed dissociation. Conditional on a content-specific internal state being instantiated in a candidate $S$, an access-poor state would have substantial reachability but weak outward observability, or strong receive and send capacities that fail to align through the same modes:
\begin{equation}
    \Lambda_c\not\approx 0,
    \qquad
    \Lambda_o\approx 0\ \text{or}\ \Aspec\ll1,
    \qquad
    Q=\Cspec\Aspec\ \text{is small}.
    \label{eq:p-without-a-phenotype}
\end{equation}
Activity from specialist systems can then enter or sustain differentiated candidate states, yet the weak observability spectrum or poor mode alignment prevents those state differences from becoming broadly consequential for memory, evaluation, flexible action, and other predeclared outputs. If an independent theory or measurement identifies the reached state as phenomenal, this receive-without-expression regime is a compact formalization of phenomenal consciousness without access. It also distinguishes that hypothesis from the absence of neural processing.

A second possibility follows from recurrent-processing theories: phenomenal content may be instantiated within local specialist circuitry and never reach the GMW. In that case, the relevant failure is low content-specific reachability into $S$, not high reachability followed by low observability. The taxonomy of conscious, preconscious, and subliminal processing can be restated in this language: a locally represented state may be potentially reachable while the current access route remains closed, whereas a weak subliminal representation may fail before a candidate GMW state is robustly reached \citep{Dehaene2006}. These alternatives are mechanistically different even when both produce no immediate report.

Report itself should not be identified with observability. The output operator can include effects on memory, attention, valuation, autonomic regulation, eye movements, and nonverbal action. No-report paradigms remove or reduce report-specific processes and have shown that some frontal activity is more closely related to introspection and action than to the perceptual transition itself \citep{Frassle2014,Tsuchiya2015}. Yet removing overt report does not demonstrate that all cognitive access is absent, and attempts to eliminate every downstream consequence risk eliminating the evidence used to classify the state as conscious \citep{PhillipsMorales2020}. In GMW terms, speech is one possible output channel within $R$, not the definition of $\Wo$.

The framework expresses the philosophical contrast as competing empirical signatures. An overflow account predicts content-specific local or candidate-state evidence together with preserved reachability and differentiation, but weak broad observability or alignment at the time of experience. A strict access account predicts that reliable phenomenality will covary with realized mediation $Q$, alignment, and routed breadth once all relevant nonverbal outputs are included. Tests must specify the candidate, input channels, and output channels before examining the data; otherwise the boundary can be adjusted post hoc to manufacture either access or its absence. GMW quantities can sharpen the comparison, but mediation alone cannot establish phenomenal status.

\subsection{What the present formulation does not yet explain}

Several features of global workspace theory remain outside the current model. The GMW evaluates whether a declared candidate mediates boundary activity, but it does not specify how competing contents enter a winner-take-most contest, why access is serial, or how one content suppresses another. The state-dependent coalition switching in the nonlinear simulations concerns which substrate instantiates mediation under a context; it is not yet a model of which representational content wins access within that substrate.

Endogenous attentional control is also treated as part of the operating context rather than derived from an explicit control policy. A fuller model would need content-specific boundary inputs, internal selection dynamics, and an objective or value signal that changes the gain of competing routes. The access--phenomenality distinction discussed above adds a further boundary: the GMW formalizes global availability and its failures, but phenomenal status cannot be read directly from the mediation signature.

These omissions leave an empirical question. Global workspace theory does not by itself determine whether conscious access depends most strongly on receive capacity, send capacity, alignment, effective dimensionality, routed breadth, or a particular conjunction. The GMW measures these components separately. Masking, attentional blink, no-report paradigms, perturbational experiments, and state transitions can test whether the predicted components change together or dissociate.

\subsection{Interpretation of the macaque ECoG analysis}

The primary contribution of this study is the formulation and synthetic validation of the GMW. The macaque analysis was used as a proof of application. It examined whether the proposed quantities can be estimated from neural recordings, whether they separate gain from organization during a change of state, and which estimation problems arise in practice. These problems included algebraic dependence induced by rereferencing, selection circularity, incomplete anatomical coverage, slow-wave amplification, flexible channel identity, dependence on temporal grain and regularization, and the locality of bipolar measurements.

In the macaque data, interaction magnitude and mode organization did not change in parallel. Short-lag predictability, $Q$, and $\Cspec$ increased, while $\Aspec$ decreased across candidate sizes. Reductions in $\Deff/k$, $G_{\mathrm{pair}}$, and $\Oorg$ were more dependent on candidate size and animal, and were clearest at $k=5$. A single read/write magnitude did not capture this pattern. Recovery also differed across components. These observations support use of the multicomponent signature, but they do not establish that the measured operator is a sufficient neural marker of consciousness.

The spatial results are broadly compatible with earlier global-workspace research without constituting anatomical confirmation. Recurrently selected sites extended across frontal and premotor, parietal and sensorimotor, temporal, and other association-related sectors. The distributed pattern is compatible with the long-range prefrontal, parietal, high-level sensory, and action-related participation emphasized in global neuronal workspace models and intracranial studies of conscious access \citep{Dehaene1998,DehaeneNaccache2001,Gaillard2009,Mashour2020}. Exact channels varied across animals and days, electrode coverage was incomplete, and the common display did not provide stereotactic registration. The state-dependent component differences were more stable than localization of a unique macaque GMW.

Two analyses would test the present interpretation more directly. Phase-randomized or autocorrelation-matched surrogates would estimate how much of the result follows from slow temporal structure, and a multianimal path-constrained analysis would estimate how much of the $q=1$ effect requires transitions between distinct channels. Further work also requires improved anatomical registration, additional anesthetics, conscious-content paradigms, direct perturbations, and broader coverage of specialist systems. The present dataset shows that the signature can be estimated and identifies several methodological constraints. It does not provide a final empirical test of GWT.

\subsection{Candidate size, scale, and inferential limits}
\label{sec:minimality}

The framework does not determine a unique candidate size. Adding a useful node can increase capacity or introduce a new mediation mode, while a redundant node can reinforce an existing route. Fixed-size search, held-out performance, size-matched nulls, stability, leave-one-out indispensability, and modal novelty should be reported separately. The same principle applies to the horizon. Each side of the boundary operator contains $L$ samples, and the $q$-shifted operator spans input--output separations from $(q+1)\Delta t$ to $(2L+q-1)\Delta t$. Distinct temporal ranges can reveal different instantiating coalitions.

Nor does the GMW select its own state variables or system boundary. A generic projection can mix a candidate with its exterior and obscure the physical meaning of boundary inputs and outputs. A canonical account of the conscious substrate would require an additional principle for choosing spatial grain, temporal grain, variables, and cut. IIT 4.0 addresses intrinsic cause--effect structure, while information-closure approaches address predictive autonomy; those criteria could, in principle, constrain the boundary before GMW mediation is evaluated \citep{Albantakis2023,Bertschinger2008,Chang2020}.

Within global workspace logic, conscious access is relational. The instantiating coalition is not a container in which consciousness resides independently of the specialist systems. Its function is to make selected states available across a larger organization. Content may remain distributed in the specialists while the GMW supplies stabilization, transformation, and routing. Establishing that a route is privileged requires perturbation or strong causal identification, because marginalizing a hidden mediator can make its contribution appear as effective recurrence in the observed remainder.

The empirical operators used here are predictive dynamics at a 25-ms interval, not direct synaptic causality. Bipolar rereferencing, filtering, hidden sources, volume conduction, and regularization shape the fitted state space. The independent montage avoids a known algebraic redundancy, and selection-safe tests reduce circularity, but neither step makes the observation model unique. The empirical conclusion is conditional on the declared ECoG representation: deep anesthesia increased short-lag mediation gain and potential capacity while reducing input--output alignment. Lower differentiated dimensionality and routed organization were more evident for larger candidate sets and varied across animals. Generalization to other anesthetics, conscious contents, species, and measurement modalities remains open.

\section{Conclusion}

The Global Mediation Workspace turns the central functional claims of global workspace theory into an explicit input--state--output problem. Reachability formalizes access from specialist systems, observability formalizes availability back to those systems, alignment tests whether the same internal modes connect the two directions, effective rank measures differentiated mediation within a limited candidate, and routed breadth quantifies the range of source--target transformations supported through the candidate. WMI remains a pragmatic search index; the theoretical object is the multicomponent signature, and the components required for conscious access must be identified empirically.

Synthetic tests show that the formulation rejects split read/write aggregates, one-sided hubs, and aligned but low-rank bottlenecks. The nonlinear extension describes an operator-level hypothesis for ignition: latent two-sided capacity becomes realized mediation when state- or amplitude-dependent alignment opens differentiated routes. The reachability--observability separation also gives a precise language for access-poor states, including the hypothesis that a state may be entered without becoming globally expressed, while leaving its phenomenal status open. The macaque ECoG application shows that the signature can distinguish increased dynamical gain from reduced input--output alignment in real recordings, while also showing that dimensionality and routed organization depend on candidate size and animal. The analysis also exposes the importance of montage construction, selection, slow temporal structure, temporal scale, and anatomical coverage. The framework permits tests of global availability without reducing it to connectivity strength or a single consciousness score.

\FloatBarrier
\clearpage
\begingroup
\raggedright
\bibliographystyle{plainnat}
\bibliography{references}
\endgroup
\clearpage

\appendix
\section{Standard linear-system identities, invariance, and shifted operators}
\label{app:linear-identities}

The results collected here are standard consequences of finite-horizon reachability, observability, Hankel-operator, and singular-value identities \citep{Kailath1980,Moore1981,Kurschner2018}. They are included to make the notation and implementation self-contained, not as claims of novelty.

\subsection{Boundary-Hankel and Gramian spectra}

The squared nonzero singular values of $\Observe_L\Reach_L$ are the nonzero eigenvalues of $\Reach_L^\top\Observe_L^\top\Observe_L\Reach_L$. The nonzero eigenvalues of $XY$ and $YX$ coincide, giving those of $\Wo\Wc$ and hence \cref{eq:eta}.

\subsection{Minimum internal path length}
\label{sec:shifted-hankel}

The unshifted operator includes the direct boundary term $C_SB_S$. When a scientific hypothesis requires at least $q$ candidate transitions, define
\begin{equation}
    \Hankel_{L,q}(S)=\Observe_L(S)A_S^q\Reach_L(S),\qquad q=0,1,2,\ldots,
    \label{eq:shifted-hankel}
\end{equation}
with blocks
\begin{equation}
    [\Hankel_{L,q}(S)]_{ij}=C_SA_S^{i+j+q}B_S.
    \label{eq:shifted-block}
\end{equation}
The choice $q=1$ removes the direct $C_SB_S$ term but can still be satisfied by same-state persistence. To require at least one transition between distinct candidate coordinates, write $A_S=D_S+E_S$ with $D_S=\operatorname{Diag}(\operatorname{diag}A_S)$ and define
\begin{equation}
    [\Hankel^{\mathrm{cross}}_{L,q}(S)]_{ij}
    =C_S\left(A_S^{m}-D_S^{m}\right)B_S,
    \qquad m=i+j+q.
    \label{eq:cross-node-hankel}
\end{equation}
This retains all words in the expansion of $(D_S+E_S)^m$ containing at least one off-diagonal transition. It differs from replacing $A_S$ by $E_S$, which removes diagonal persistence at every step.

With
\begin{equation}
    \Wo^{(q)}=(A_S^\top)^q\Wo A_S^q,
    \label{eq:shifted-wo}
\end{equation}
the shifted singular values are
\begin{equation}
    \eta_i^{(q)}(S;L)=\sqrt{\lambda_i\!\left(\Wo^{(q)}\Wc\right)}.
    \label{eq:shifted-eta}
\end{equation}

\subsection{Internal similarity invariance}

For an invertible internal coordinate change $z'=Tz$,
\begin{equation}
    A'_S=TA_ST^{-1},\qquad B'_S=TB_S,\qquad C'_S=C_ST^{-1}.
\end{equation}
Each Markov block is unchanged because $C'_S(A'_S)^\ell B'_S=C_SA_S^\ell B_S$. Consequently $\Hankel'_L=\Hankel_L$. The total mediation spectrum is invariant to internal reparameterization, although the capacity--alignment decomposition depends on the declared state metric under nonorthogonal transformations.

\subsection{Cross-basis and capacity bounds}

Using \cref{eq:eigendecomp},
\begin{equation}
    \Wo^{1/2}\Wc^{1/2}
    =U_o\Lambda_o^{1/2}(U_o^\top U_c)\Lambda_c^{1/2}U_c^\top,
\end{equation}
so orthogonal invariance of singular values gives \cref{eq:crossfactor}. Because $M=U_o^\top U_c$ is orthogonal, the von Neumann--Schatten rearrangement inequality yields
\begin{equation}
    Q_L=\|\Lambda_o^{1/2}M\Lambda_c^{1/2}\|_*
    \leq\sum_i\sqrt{\lambda_i(\Wc)\lambda_i(\Wo)}=\Cspec,
\end{equation}
which establishes \cref{eq:decomp}.

\subsection{Zero mediation and subspace orthogonality}

Let $\mathcal C_L=\im(\Reach_L)$ and $\mathcal O_L=\im(\Observe_L^\top)$. Then
\begin{equation}
    \Hankel_L=0
    \quad\Longleftrightarrow\quad
    \im(\Reach_L)\subseteq\ker(\Observe_L)
    \quad\Longleftrightarrow\quad
    \mathcal C_L\perp\mathcal O_L.
\end{equation}
A graph with no directed path from the support of $B_S$ to states read by $C_S$ is a sufficient structural case. Graph connectivity is not sufficient for nonzero mediation because signed cancellations or orthogonal subspaces can still eliminate transfer.

\section{Additional alignment and mode diagnostics}
\label{app:alignment-details}

\subsection{Principal angles}

For dominant controllability and observability eigenvector matrices $U_c^{(d)}$ and $U_o^{(d)}$, principal-angle cosines are
\begin{equation}
    \alpha_j^{(d)}=\sigma_j\!\left[(U_o^{(d)})^\top U_c^{(d)}\right]=\cos\theta_j.
    \label{eq:principal}
\end{equation}
They are useful descriptive summaries but do not replace the energy-weighted alignment $\Aspec$.

\subsection{Metric dependence}

The total boundary operator and its singular spectrum are invariant under an internal similarity transformation. The separate factors $\Cspec$, $\Aspec$, and Euclidean principal angles depend on the state metric under a nonorthogonal transformation because $\Wc$ and $\Wo$ transform by different congruences. Empirical reports must therefore state the normalization or covariance metric used to define equal state energy.

\subsection{Compact balancing and node--mode roles}

Factor $\Wc=X_cX_c^\top$ and $\Wo=X_oX_o^\top$, and compute the compact singular-value decomposition
\begin{equation}
    X_o^\top X_c=U_r\Sigma_rV_r^\top,
    \qquad \Sigma_r=\operatorname{diag}(\eta_1,\ldots,\eta_r),
    \label{eq:compact-svd}
\end{equation}
where $r=\rank(\Hankel_L)$. Define
\begin{equation}
    T_r=X_cV_r\Sigma_r^{-1/2},\qquad
    S_r=\Sigma_r^{-1/2}U_r^\top X_o^\top.
    \label{eq:balanceT}
\end{equation}
Then
\begin{equation}
    S_rT_r=I_r,\qquad
    S_r\Wc S_r^\top=\Sigma_r,\qquad
    T_r^\top\Wo T_r=\Sigma_r.
    \label{eq:balanced}
\end{equation}
A reciprocal-scale-invariant node--mode participation diagnostic is
\begin{equation}
    p_{ji}=\frac{|(T_r)_{ji}(S_r)_{ij}|}
    {\sum_{q\in S}|(T_r)_{qi}(S_r)_{iq}|}.
    \label{eq:participation-balanced}
\end{equation}
Exactly degenerate modes should be interpreted as a joint subspace because rotations within that subspace are not unique.

\subsection{Subspace-valued candidates}

A distributed GMW could be represented by a rank-$k$ projector rather than an axis-aligned node subset. Optimization over a Grassmann manifold is mathematically possible \citepapp{Edelman1998}, but a generic projector mixes the candidate with its exterior and makes the physical boundary ambiguous. We therefore use balanced subspaces after identifying a node set and leave intrinsic subspace-boundary selection to future work.

\section{Normalization, horizon, recurrence, and candidate size}
\label{app:parameter-choice}

\subsection{State metric and covariance weighting}

The singular values inherit units from the state and boundary coordinates. Empirical analyses should declare a whole-system normalization. Covariance-weighted channels can be defined by
\begin{equation}
    \widetilde B_S=B_S\Sigma_u^{1/2},\qquad
    \widetilde C_S=\Sigma_y^{-1/2}C_S,
\end{equation}
with regularized inverse square roots when necessary. The alignment decomposition must use the same metric.

\subsection{Finite horizon}

The horizon parameter $L$ specifies the number of past input samples and the number of future output samples used by the finite operator. Under the indexing in \cref{eq:markov}, the unshifted blocks represent input--output separations from $\Delta t$ to $(2L-1)\Delta t$. With a required internal shift $q$, the range is $(q+1)\Delta t$ to $(2L+q-1)\Delta t$. Thus $L\Delta t$ describes the nominal extent of each history window, not the largest mediated lag. A defensible choice uses the shortest horizon after which the candidate ranking and signature stabilize while remaining within the timescale on which the fitted dynamics are credible. Spectral decay supplies a complementary diagnostic. Horizon sensitivity should distinguish reevaluation of a fixed candidate from candidate reselection.

The synthetic linear benchmark uses $L=10$, the nonlinear activation benchmark $L=6$, and the macaque analysis $L=4$ at a 25-ms prediction lag. These choices specify analysis-specific temporal ranges; they do not define a universal GMW timescale.

\subsection{Candidate size and minimality}

WMI is conditional on size. Useful additions can increase capacity, routed breadth, or mode count after a compact mechanism is already present. Candidate-size inference should combine held-out prediction, size-matched nulls, stability, complexity penalties, leave-one-out indispensability, and modal novelty. The present framework does not supply a canonical rule for choosing $k$.

\begin{figure}[t]
    \centering
    \includegraphics[width=\textwidth]{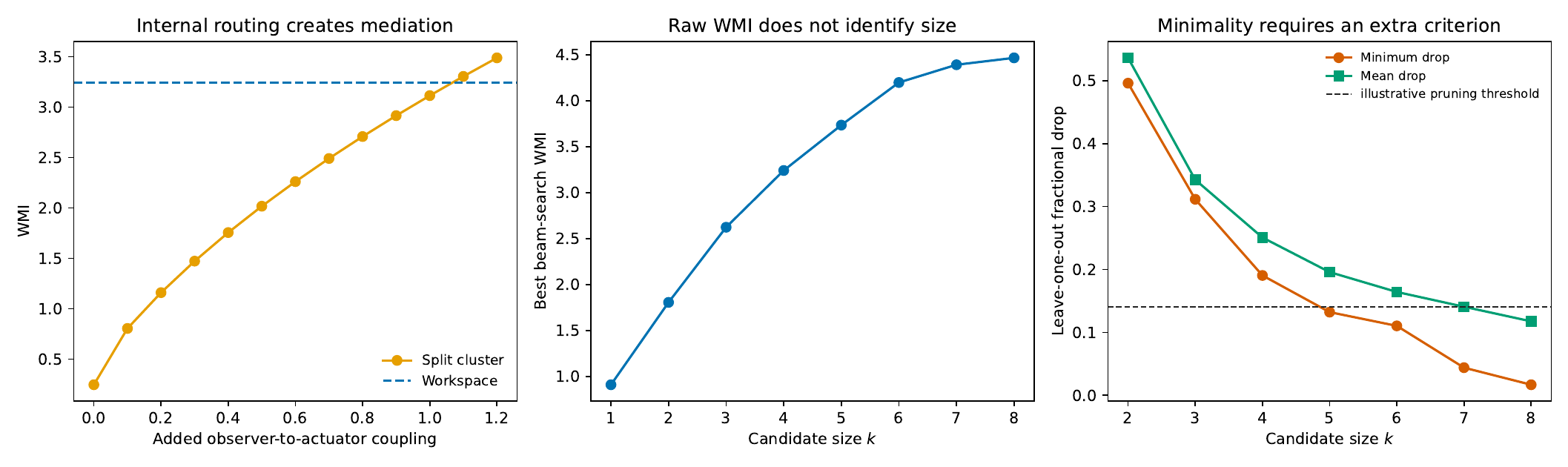}
    \caption{\textbf{Internal routing and candidate size.} Adding the missing route turns the split decoy into a mediator. Raw WMI can continue to reward useful additions beyond the planted four-node set, so fixed-size identification and minimality are separate problems.}
    \label{fig:minimality}
\end{figure}

\subsection{Module partition and scalarization}

The routed-breadth term $G_{\mathrm{pair}}$ is conditional on the partition of the remainder into specialist modules. A coarse partition can merge functionally distinct routes, whereas a fine partition can distribute the same mediated energy across many module pairs. The current framework does not determine a unique module granularity. Empirical analyses should define the partition before state comparison and, when routed breadth is central to the conclusion, examine a small set of plausible alternatives. This is an unresolved problem of scale selection, similar to the choice of candidate size.

The multiplicative WMI in \cref{eq:wmi} is a pragmatic fixed-size search rule. It is useful because a near-zero value in any component lowers the combined score, but the product form and equal implicit weighting are not unique. Weighted geometric means, log-additive scores, or Pareto-based selection could produce different rankings. For this reason, the four-component signature remains the primary result and WMI should be used only within a prespecified analysis design. Determining which scalarization, if any, best predicts conscious access is left open.

\section{Computational evaluation without forming the large Hankel matrix}

Although $\Hankel_L$ has dimension $L|R|\times L|R|$, at most $k$ singular values are nonzero. We compute them from the $k\times k$ matrix $\Wo^{1/2}\Wc\Wo^{1/2}$. Module-pair energies are evaluated as $\tr(\Wo{}_b\Wc{}_a)$. For fixed small $k$, the dominant per-candidate cost is forming boundary covariance terms and propagating $k\times k$ matrices over $L$ steps. Parallel compiled kernels were used for exact enumeration.

\section{Alternative trace-normalized alignment decomposition}
\label{app:fidelity-alignment}

For completeness, an alternative normalization separates total receive/send energy from trace-normalized overlap:
\begin{equation}
    \Ctot=\sqrt{\tr(\Wc)\tr(\Wo)},\qquad
    \AF=\frac{\normnuc{\Wo^{1/2}\Wc^{1/2}}}{\sqrt{\tr(\Wc)\tr(\Wo)}}.
\end{equation}
Then $Q_L=\Ctot\AF$ and $0\leq\AF\leq1$ by $\|XY\|_*\leq\|X\|_F\|Y\|_F$. If $\rho_c=\Wc/\tr\Wc$ and $\rho_o=\Wo/\tr\Wo$, $\AF=\tr[(\rho_o^{1/2}\rho_c\rho_o^{1/2})^{1/2}]$, the root fidelity associated with Bures--Wasserstein geometry \citepapp{BhatiaJainLim2019}. The spectral normalization $\Aspec$ is used in the main text because it directly compares realized mediation with the maximum allowed by the two Gramian spectra.

\section{Exact metric definitions used in the conventional comparison}

Graph baselines used $|A|$ and excluded diagonal self-dynamics. Directed betweenness used source-to-target edges corresponding to $A_{\mathrm{target},\mathrm{source}}$. Participation used combined incoming and outgoing strength to each specialist module. Communicability used the matrix exponential of the absolute directed matrix after spectral-radius normalization. Average and modal controllability used the eigendecomposition of the symmetrized absolute matrix, rescaled below unit spectral radius:
\begin{align}
    \phi_i^{\mathrm{avg}}&=\sum_j\frac{v_{ij}^2}{1-\lambda_j^2},\\
    \phi_i^{\mathrm{modal}}&=\sum_j(1-\lambda_j^2)v_{ij}^2.
\end{align}
These conventions are transparent instantiations rather than uniquely privileged definitions.

\section{Mathematical and numerical verification}

The cross-basis factorization reproduced direct boundary-Hankel singular values with maximum absolute error below $10^{-14}$ across the prespecified linear candidates. Direct Hankel-versus-Gramian comparison differed by at most $2.1\times10^{-12}$, with the largest discrepancy confined to nearly zero singular values of the rank-one hub. Compact balancing residuals in $S_rT_r-I_r$, $S_r\Wc S_r^\top-\Sigma_r$, and $T_r^\top\Wo T_r-\Sigma_r$ were below $1.9\times10^{-12}$. A separate rank-deficient four-state test yielded a two-mode compact realization with residuals below $1.2\times10^{-15}$. The nonlinear differential factorization was independently verified on random linear time-varying systems by comparing finite differences of the nonlinear boundary map with $\Observe^+\Reach^-$ over decreasing perturbation amplitudes.

\section{Additional nonlinear theory and simulation details}
\label{app:nonlinear-details}

\subsection{Finite-amplitude ignition and saturation}

The activity-gated four-state system used 64 perturbation directions and 18 propagation steps. The gate slope was 14 and its threshold was 0.58. The secant strength $Q^{\mathrm{FA}}$ peaked at input amplitude approximately $1.053$ with value $1.005$ and effective rank $1.994$; at amplitude 2 it decreased to $0.744$ because of saturation. The peak therefore reflects a finite-amplitude operating regime, not a monotonic consciousness scale.

\begin{figure}[t]
    \centering
    \includegraphics[width=\textwidth]{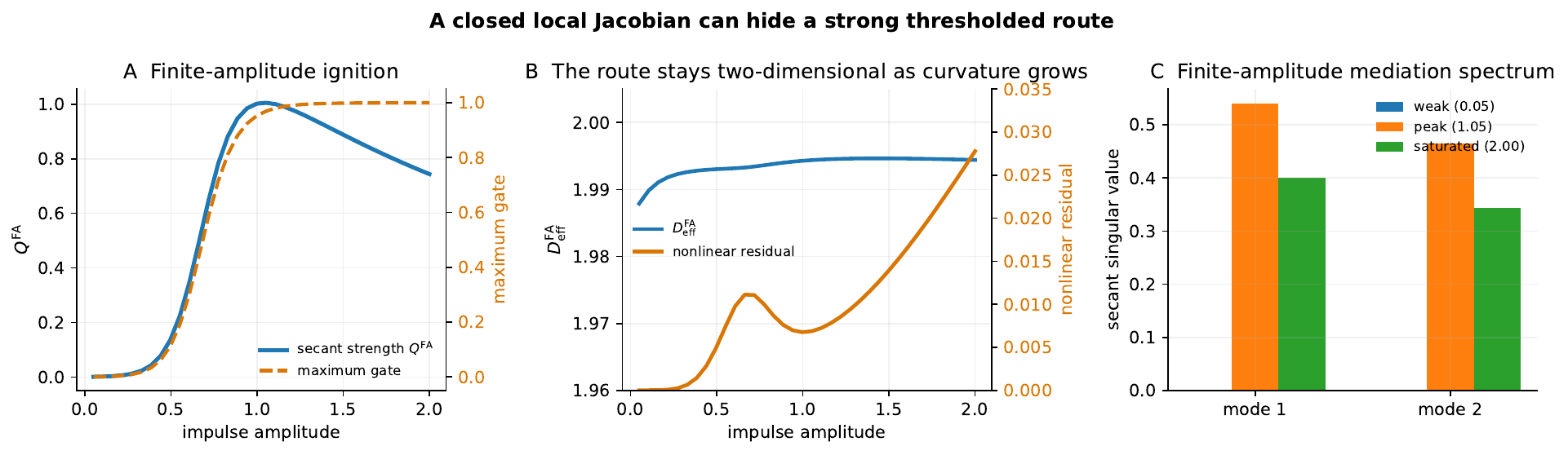}
    \caption{\textbf{Finite-amplitude mediation.} A route that is weak under infinitesimal perturbation opens at intermediate amplitude and then saturates. The secant profile is conditional on the reference trajectory, amplitude, and perturbation ensemble.}
    \label{fig:nl-finite-appendix}
\end{figure}

\subsection{Hysteresis and higher-order mediation}

In the bistable context example, opening and closing transitions occurred at drives $+0.305$ and $-0.305$. At zero drive, the closed branch had WMI $0.0019$ and the open branch WMI $1.249$. This demonstrates possible dynamical hysteresis in the simulation; it should not be conflated with the macaque recovery result, where anesthetic concentration was not matched.

A multiplicative two-source example had zero first-order norm for each source separately and mixed second-derivative norm $1.637$. This shows that a route can exist only in a joint source configuration. The quantity is not PID synergy and is not folded into canonical WMI.

\begin{figure}[t]
    \centering
    \begin{subfigure}[t]{0.48\textwidth}
        \centering
        \includegraphics[width=\textwidth]{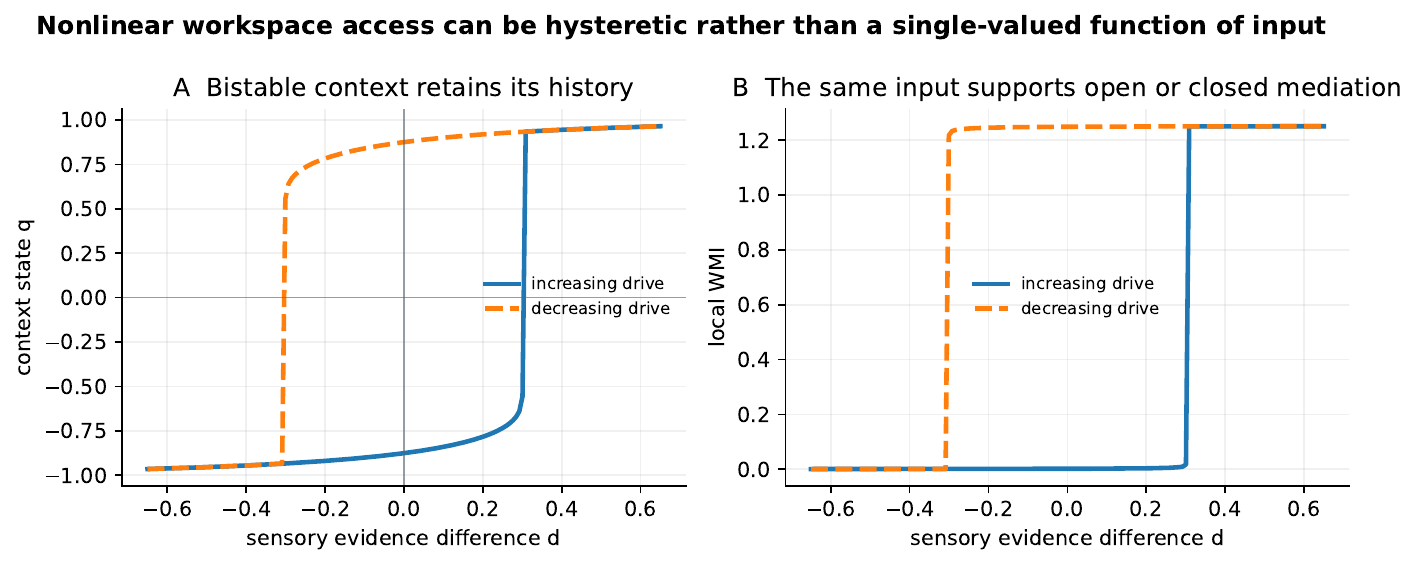}
        \caption{Bistable context and simulated hysteresis.}
    \end{subfigure}\hfill
    \begin{subfigure}[t]{0.48\textwidth}
        \centering
        \includegraphics[width=\textwidth]{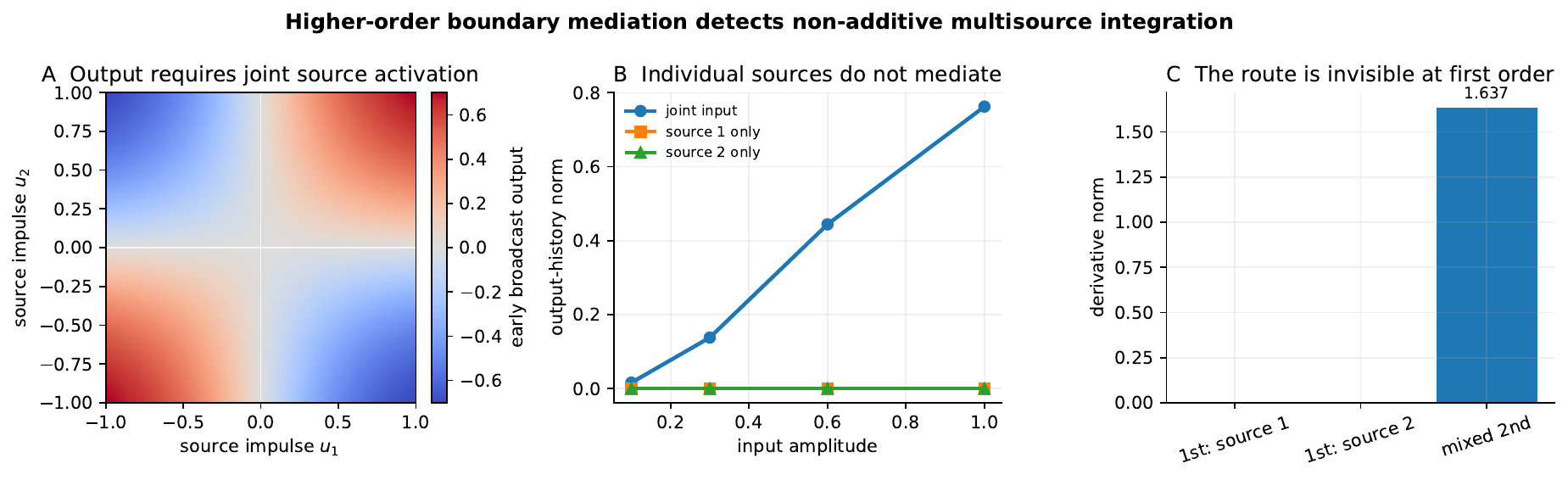}
        \caption{A route visible only in the mixed second-order response.}
    \end{subfigure}
    \caption{\textbf{Nonlinear effects beyond the local spectrum.} Hysteresis is a property of a bistable trajectory, whereas higher-order mediation measures non-additive source interactions.}
    \label{fig:nl-extra}
\end{figure}

\subsection{Activation benchmark design and inferential levels}

The compact activation benchmark contained four five-node specialist modules, a planted four-node mediator, a receiver-only set, a broadcaster-only set, a dense approximately rank-one hub, and a split input/output decoy. The spectral radius of the preactivation linear matrix was $0.88$, with $L=6$ and $q=1$. All 58,905 four-node candidates were scored. The 64-node confirmation was an independent dimension-matched realization with 635,376 candidates, not the identical matrix used in the linear benchmark.

Three operators were compared: a trajectory mean Jacobian, a one-step central secant at $\epsilon=0.1$, and a passive ridge VAR(1). The mean Jacobian is a stationary effective-operator stress test, not the exact trajectory-conditioned LTV Hankel product. The secant is likewise a one-step effective transition rather than the full finite-horizon nonlinear secant map. These approximations were used to test fixed-topology core recovery under standard activations.

At tanh gain $g=0.5$, passive VAR rank improved from 36,319 to 1 as the number of transitions increased from 500 to 32,000, while oracle operators already ranked the core first. At $g=1.4$ and $1.6$, oracle ranks also deteriorated, indicating that the active nonlinear dynamics had changed the functional winner. For ReLU, mean-Jacobian/central-secant/VAR ranks at bias $0.05$ were $2/1/245$; at bias $0.15$ all were 1. The hard-threshold pathwise Jacobian was identically uninformative, whereas a noise-averaged operator and logistic models recovered the core at intermediate switching noise. These results motivate reporting structural, functional, and statistical recovery separately.

\begin{figure}[t]
    \centering
    \includegraphics[width=\textwidth]{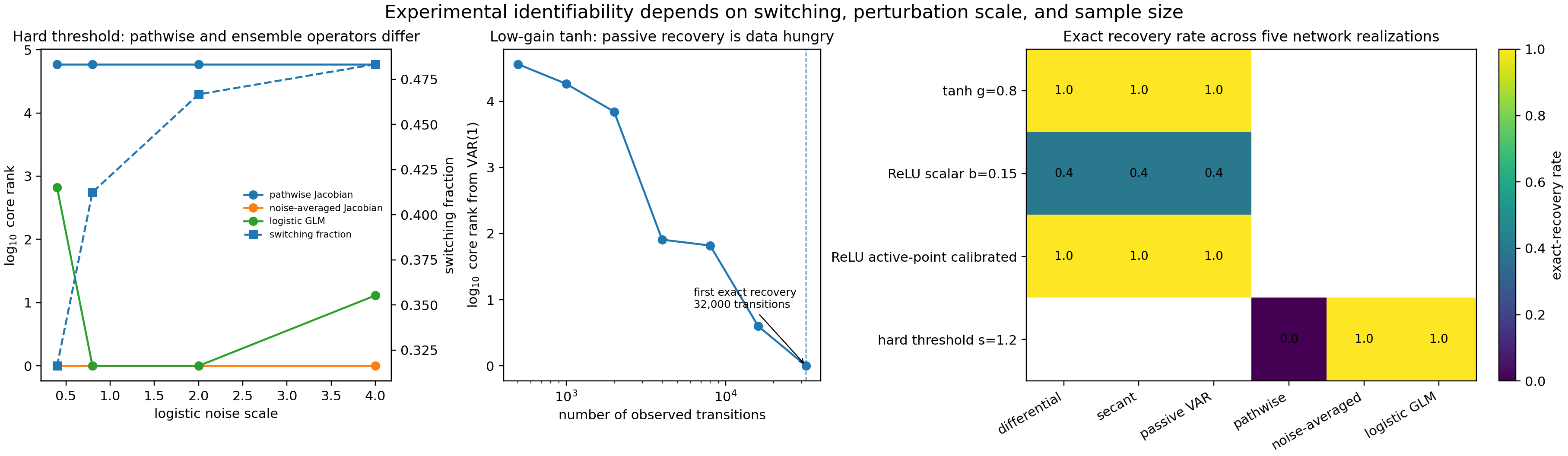}
    \caption{\textbf{Additional activation-function controls.} Passive recovery improves with sample size, and hard-threshold recovery depends on an ensemble with informative switching. Numerical thresholds are specific to the declared generative model and input distribution.}
    \label{fig:nl-activation-extra}
\end{figure}

\section{Macaque ECoG: detailed methods}
\label{app:macaque-methods}

\subsection{Sessions, states, and independent bipolar montage}

The final analysis included 11 ketamine--medetomidine experiment days from four macaques. Awake periods were divided into eyes-open and eyes-closed blocks where available. Deep-anesthesia epochs were defined from experiment event markers with guard intervals around state boundaries. Nine days also contained recovery segments. Raw recordings were sampled at 1 kHz and preprocessing retained state boundaries.

The final multianimal analysis used no-contact-reuse bipolar pairs constructed from the animal-specific electrode layouts. If $x_i(t)$ is the monopolar voltage at contact $i$, each bipolar variable was $b_{ij}(t)=x_i(t)-x_j(t)$, and each contact occurred in at most one selected pair. Two animals retained 63 map-grounded variables. Deterministic geometry-aware matching on the other two animal-specific maps yielded 64 independent pairs in each animal, using all 128 contacts exactly once. Every incidence matrix was full row rank. This construction ensures algebraic independence of the bipolar coordinates and provides direct physical endpoints for each variable.

The earlier C2 all-neighbor montage contained 267 local differences generated from 128 contacts and had rank 126. A direct 267-variable VAR produced allocation artifacts among channels sharing contacts. Row-space projection repaired that historical analysis, but the independent montage was adopted for the final multianimal test because it provides an interpretable full-rank coordinate system without a redundant inverse representation.

\subsection{Preprocessing and state-space identification}

Monopolar data were detrended, line-noise filtered, band limited, resampled to 200 Hz, and converted to bipolar variables. Candidate discovery and state evaluation used a 25-ms predictive interval. Scaling was estimated from awake data and then held fixed when applying the model to deep and recovery states. Ridge regularization was selected within awake data; stability handling and all conventions were held fixed across states. The model was
\begin{equation}
    x_{t+5}=Ax_t+\epsilon_t
\end{equation}
for data at 200 Hz. The primary mediation analysis used $L=4$ and $q=1$.

Candidate sets sharing a physical contact were impossible by construction in the primary montage. Search was performed separately for $k=3,4,5$. Candidates were ranked by canonical WMI at fixed size, but all inferential comparisons report the full signature. The recurrence shift requires one internal state transition but does not prove a transition between distinct candidate channels.

\subsection{Selection separation and hierarchical summaries}

Within-day awake cross-fitting selected candidates on one awake partition and evaluated them on held-out awake and deep partitions. Across candidate sizes and folds this produced 132 comparisons. Same-animal leave-one-day-out analysis pooled the remaining days of the same animal to select a candidate and evaluated the held-out day, yielding 33 day-by-size evaluations. Block-wise LODO trajectories contributed 954 repeated estimates but were used for temporal characterization, not as independent biological observations.

For the full-state summaries, ratios were aggregated geometrically across days for each animal and candidate size before obtaining an animal-balanced geometric mean. Descriptive intervals resampled animals as clusters. For the normalized blockwise display, every state was divided into six equal-duration blocks. Each metric was first divided by the geometric mean across the 12 awake blocks from the same day and candidate size. Values were then averaged geometrically across $k=3$--5 and days within animal, followed by a second animal-level rescaling that set the combined awake eyes-open and eyes-closed geometric mean to one. The equal-animal geometric mean and log-scale standard error were then computed across animals. Gaps between states were retained because the blocks are ordered within states but do not form a continuous pharmacokinetic time series. With only four animals, these summaries show heterogeneity and are not population-level time-course estimates. The unnormalized blockwise display uses the original metric units and an across-animal arithmetic mean.

\begin{table}[p]
\centering
\caption{Same-animal LODO deep/awake ratios after the corrected geometry-aware rerun. Brackets give descriptive animal-cluster bootstrap intervals.}
\label{tab:macaque-lodo-full}
\scriptsize
\begin{tabular}{lccc}
\toprule
Metric & $k=3$ & $k=4$ & $k=5$ \\
\midrule
$Q$ & 2.467 [1.659, 3.797] & 2.134 [1.458, 3.571] & 2.346 [1.657, 3.639] \\
$\Cspec$ & 2.670 [1.735, 4.480] & 2.355 [1.585, 4.253] & 2.642 [1.851, 4.635] \\
$\Aspec$ & 0.924 [0.848, 0.991] & 0.906 [0.838, 0.980] & 0.888 [0.799, 0.988] \\
$\Deff/k$ & 0.941 [0.818, 1.068] & 0.952 [0.893, 1.025] & 0.906 [0.805, 0.992] \\
$G_{\mathrm{pair}}$ & 0.986 [0.668, 1.585] & 0.976 [0.821, 1.160] & 0.831 [0.691, 1.061] \\
$\Oorg$ & 0.927 [0.547, 1.641] & 0.930 [0.789, 1.095] & 0.753 [0.557, 1.055] \\
Top-mode share & 1.079 [0.880, 1.324] & 1.090 [0.930, 1.264] & 1.208 [0.998, 1.464] \\
Raw WMI & 2.287 [1.544, 3.388] & 1.984 [1.487, 2.885] & 1.768 [1.044, 2.927] \\
\bottomrule
\end{tabular}
\end{table}

\subsection{Held-out selection and cross-day stability}

Within-day cross-fitting reproduced the prespecified component directions across 132 held-out comparisons. $Q$ increased in 127 comparisons and $\Cspec$ in 128. Decreases occurred in 88 comparisons for $\Aspec$, 96 for $\Deff/k$, 98 for $G_{\mathrm{pair}}$, and 102 for $\Oorg$; top-mode share increased in 100 comparisons and raw WMI in 112. Exact candidate identity was less stable. Mean pairwise Jaccard overlap ranged from 0.14 to 0.78 across animals and candidate sizes and was generally higher at $k=5$. The state signature transferred more consistently than a fixed bipolar-channel set.

\begin{figure}[t]
    \centering
    \includegraphics[width=0.94\textwidth]{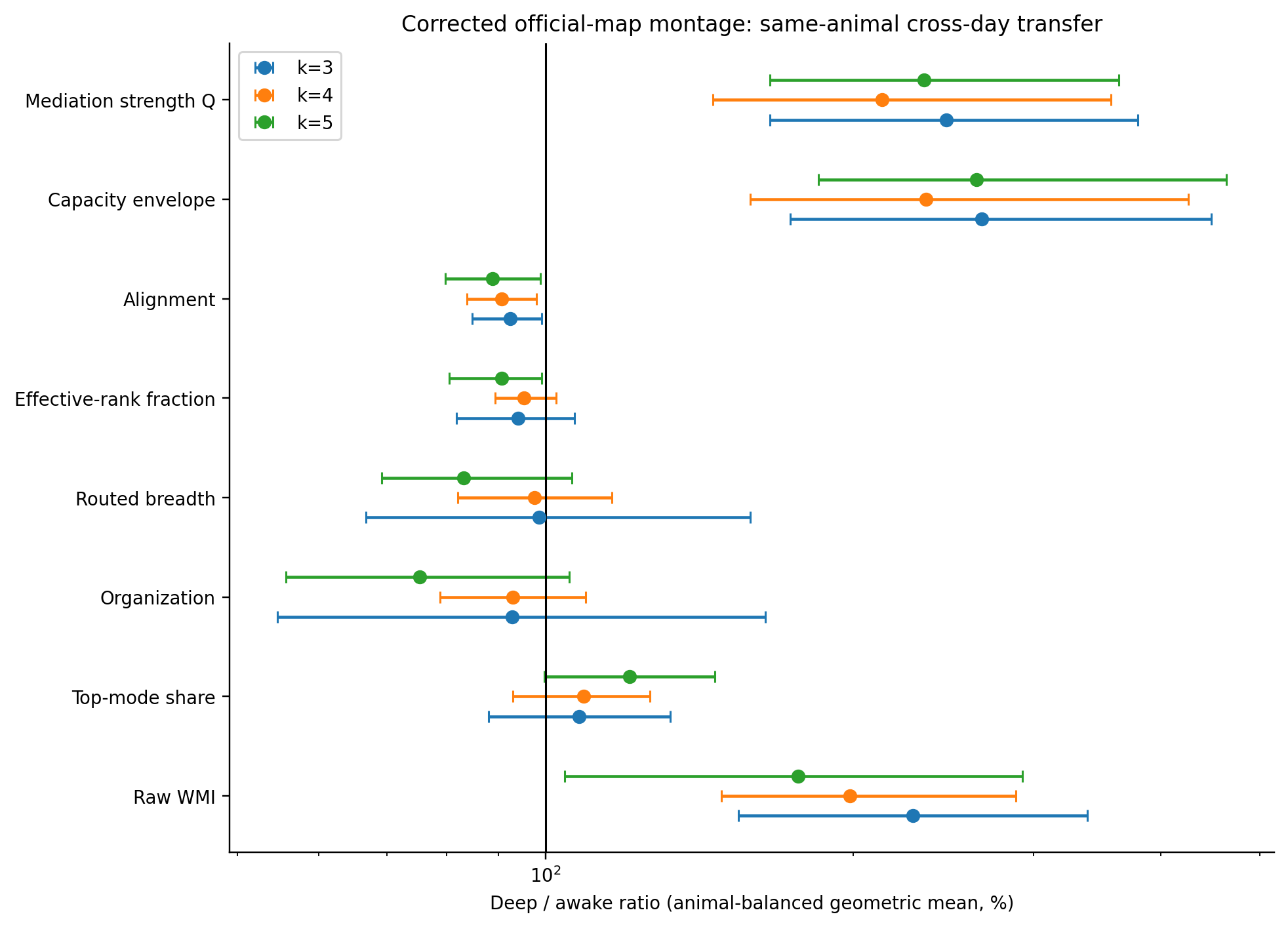}
    \caption{\textbf{Corrected same-animal cross-day transfer.} Deep/awake ratios are shown for candidates selected from the other days of the same animal. Points are equal-animal geometric means; intervals are descriptive animal-cluster bootstrap intervals. $Q$ and $\Cspec$ increase and $\Aspec$ decreases at all candidate sizes. Reductions in effective-rank fraction, routed breadth, and gain-free organization are most evident at $k=5$.}
    \label{fig:macaque-lodo}
\end{figure}

\begin{figure}[t]
    \centering
    \includegraphics[width=0.92\textwidth]{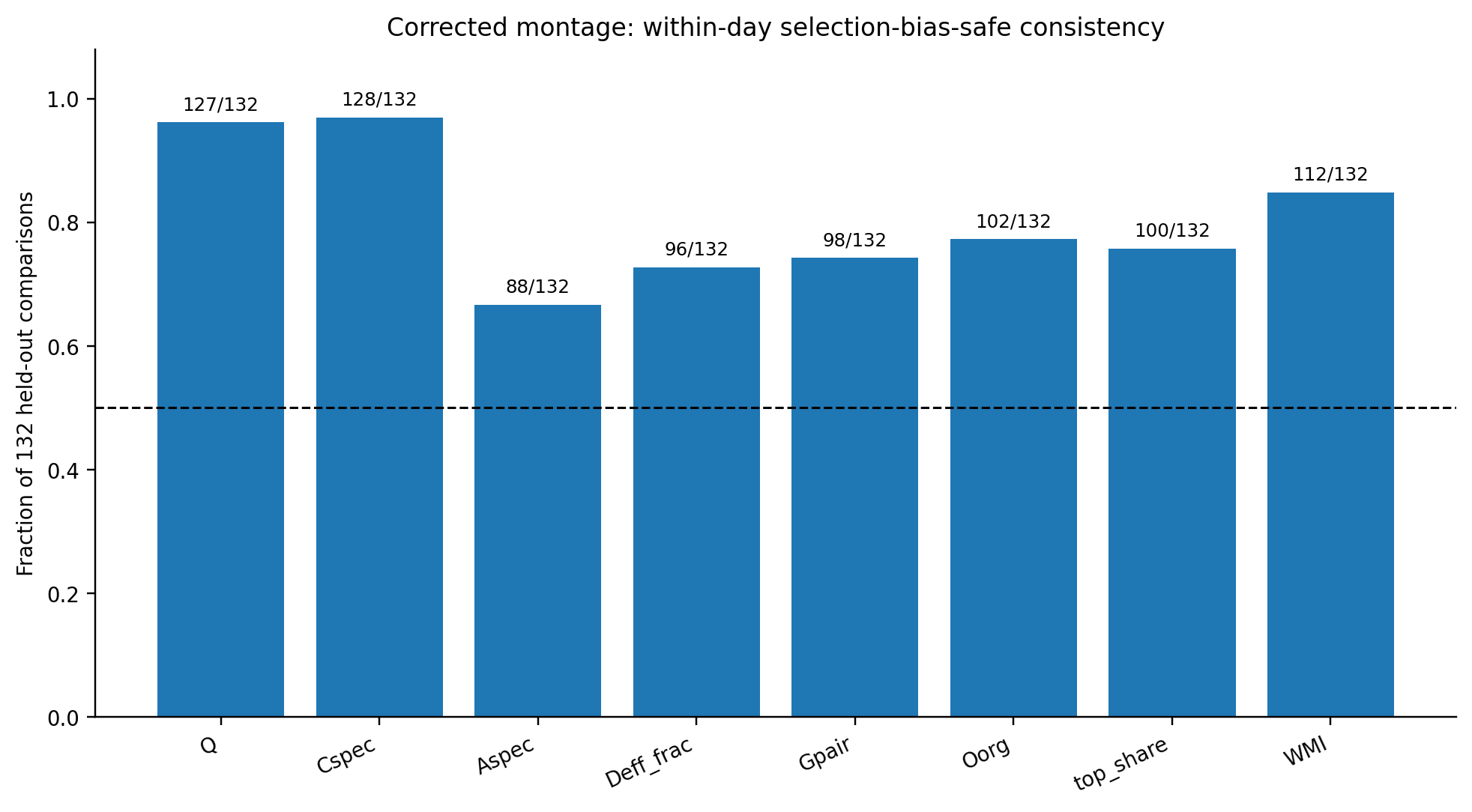}
    \caption{\textbf{Held-out directional consistency after corrected rematching.} Bars show the fraction of 132 within-day awake cross-fit comparisons with the prespecified state-effect direction. Candidate selection and state evaluation use disjoint awake data.}
    \label{fig:macaque-crossfit}
\end{figure}

\subsection{Predictive and spectral state diagnostics}

For each day, predictive $R^2$, RMS amplitude, delta fraction, autocorrelation, mean absolute correlation, leading-PC variance, and covariance effective rank were computed in awake and deep states. Deep/awake median ratios for RMS amplitude, delta fraction, and 25-ms autocorrelation were close to two, while covariance effective rank was below one on 10 of 11 days. These diagnostics characterize the observation regime in which the mediation signature was estimated.

\begin{figure}[t]
    \centering
    \begin{subfigure}[t]{0.48\textwidth}
        \centering
        \includegraphics[width=\textwidth]{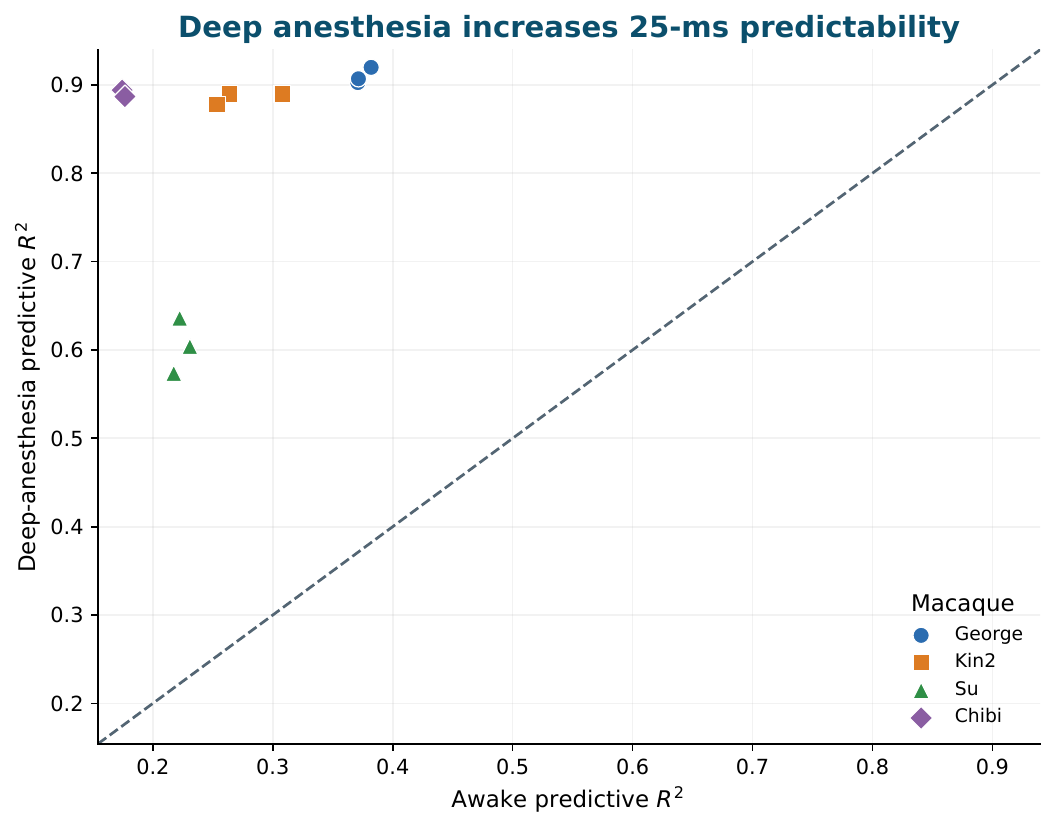}
        \caption{Held-out short-lag prediction across all days; animal identity is shown by marker and color.}
    \end{subfigure}\hfill
    \begin{subfigure}[t]{0.48\textwidth}
        \centering
        \includegraphics[width=\textwidth]{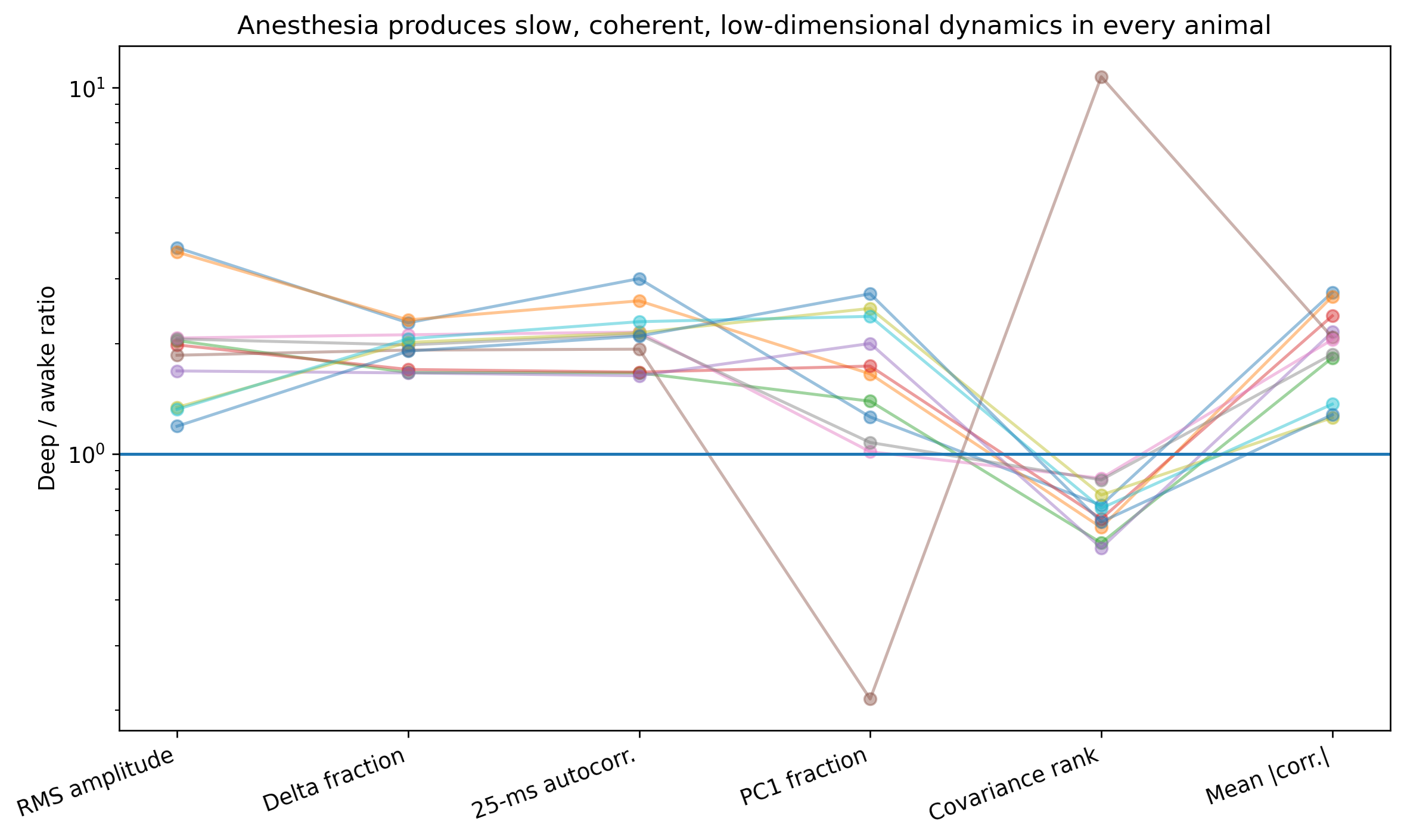}
        \caption{Slow-wave, coherence, and dimensionality diagnostics.}
    \end{subfigure}
    \caption{\textbf{Whole-network state diagnostics.} Deep anesthesia produces highly predictable, slow, correlated, low-dimensional activity.}
    \label{fig:macaque-qc}
\end{figure}

\section{Macaque ECoG: robustness, recovery, and nonlinear analysis}
\label{app:macaque-robustness}

\subsection{Slow-wave and amplitude controls}

State-wise variance normalization, restriction above 4 Hz, delta-only restriction, and state-specific leading-PC removal were applied to the LODO candidates. The corrected-montage analysis reproduced the increase in $Q$ under every transformation. Animal-balanced deep/awake $Q$ ratios for $k=3,4,5$ were $2.51/2.11/2.34$ in broadband data, $2.03/1.68/1.85$ after state-wise variance normalization, $2.30/2.35/2.51$ above 4 Hz, $4.04/4.03/3.97$ in the 0.5--4-Hz band, and $2.84/2.32/2.59$ after removal of the leading state-specific principal component.

$\Oorg$ remained below one in all displayed transform-by-size summaries. Broadband ratios were $0.89/0.93/0.74$, and the strongest reductions occurred for larger candidates and in the above-4-Hz analysis. These controls indicate that uniform amplitude scaling, the sub-4-Hz band alone, or one leading component is insufficient to explain the gain increase. They do not replace phase-randomized or autocorrelation-matched surrogates.

\begin{figure}[t]
    \centering
    \includegraphics[width=0.96\textwidth]{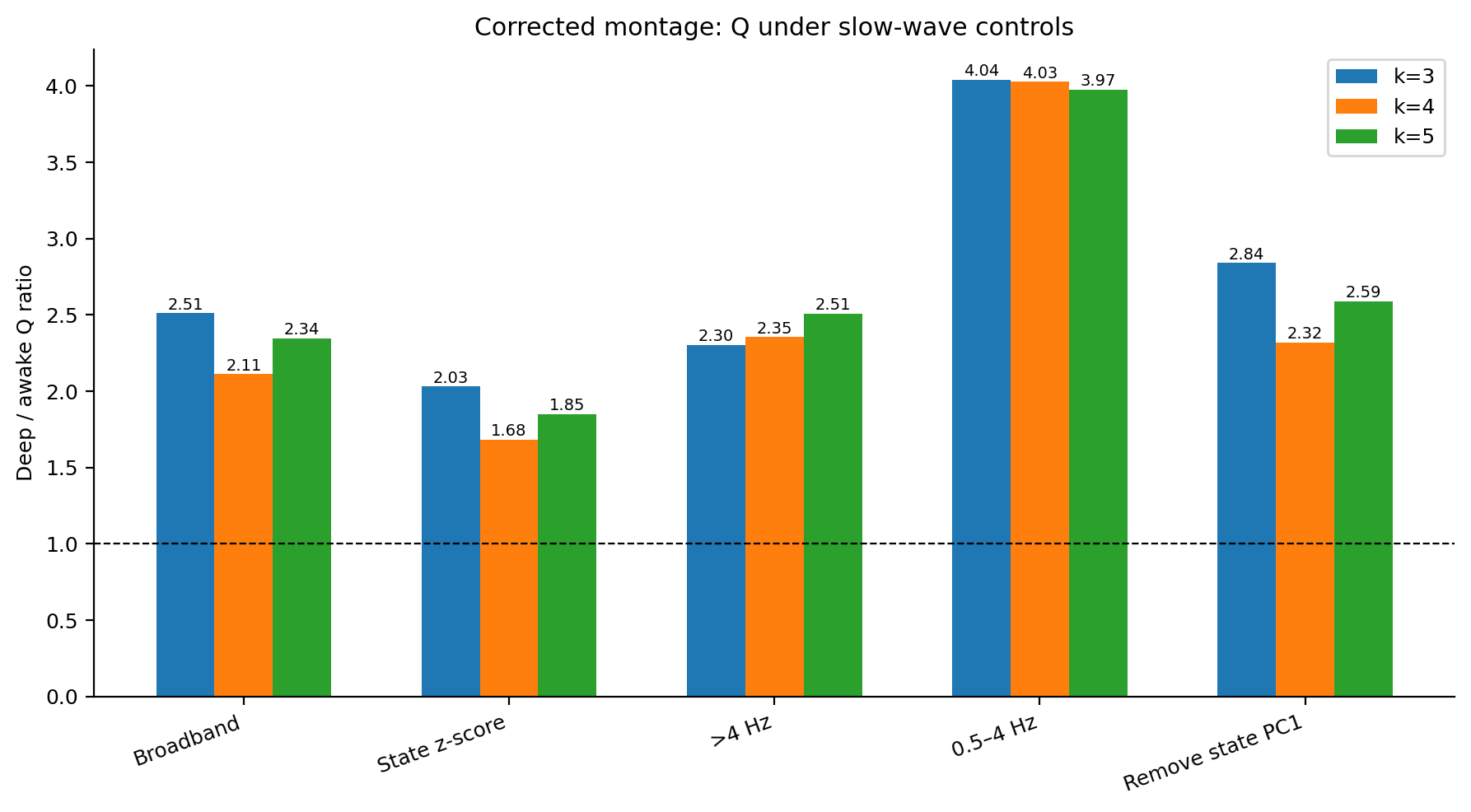}
    \caption{\textbf{Mediation strength under corrected slow-wave controls.} Animal-balanced deep/awake $Q$ ratios remain above one after state-wise variance normalization, restriction above 4 Hz, delta-only restriction, and removal of the leading state-specific principal component.}
    \label{fig:macaque-slow-controls}
\end{figure}

\begin{figure}[t]
    \centering
    \includegraphics[width=0.96\textwidth]{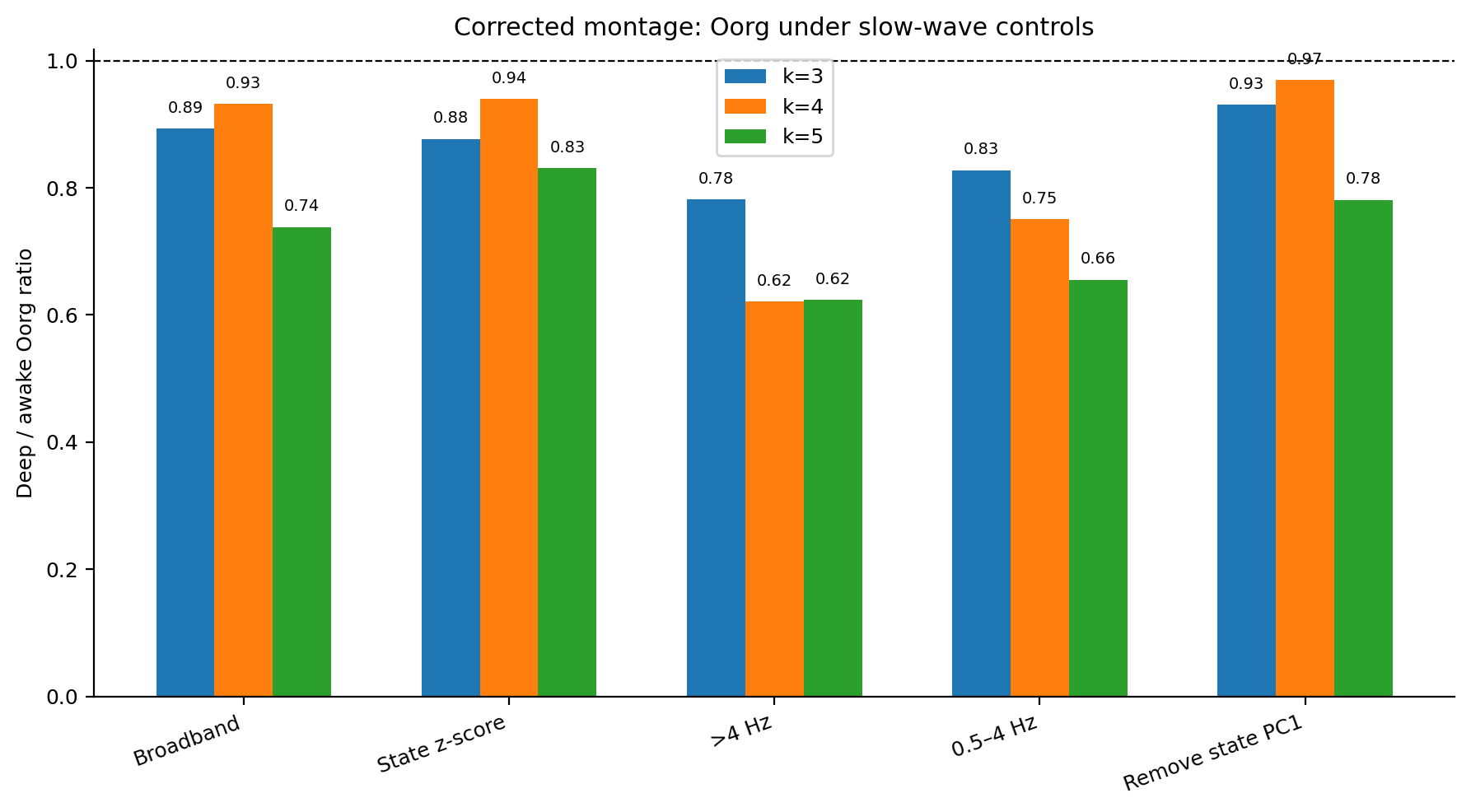}
    \caption{\textbf{Gain-free organization under corrected slow-wave controls.} $\Oorg$ remains below one across the displayed transformations and candidate sizes, with larger reductions at larger $k$ and above 4 Hz. The summaries are descriptive.}
    \label{fig:macaque-oorg-controls}
\end{figure}

\subsection{Montage locality and regional sensitivity}

The corrected geometry-aware rerun was completed for all three days in each of the two animals that required rematching. Both corrected montages contained 64 independent pairs, used every contact exactly once, and had full-row-rank incidence matrices. The group-level increase in $Q$ and $\Cspec$ and decrease in $\Aspec$ survived rematching, although exact effect sizes, selected channels, and some animal-by-size component directions changed. For example, the $k=4$ $Q$ ratio changed from $3.70$ to $2.08$ in one animal and from $1.61$ to $1.34$ in the other. The corresponding $\Oorg$ ratios changed from $0.71$ to $0.86$ and from $0.45$ to $1.03$. The robust claim concerns the multicomponent gain--alignment pattern, not invariance of every score or location.

\begin{figure}[p]
    \centering
    \includegraphics[width=0.96\textwidth]{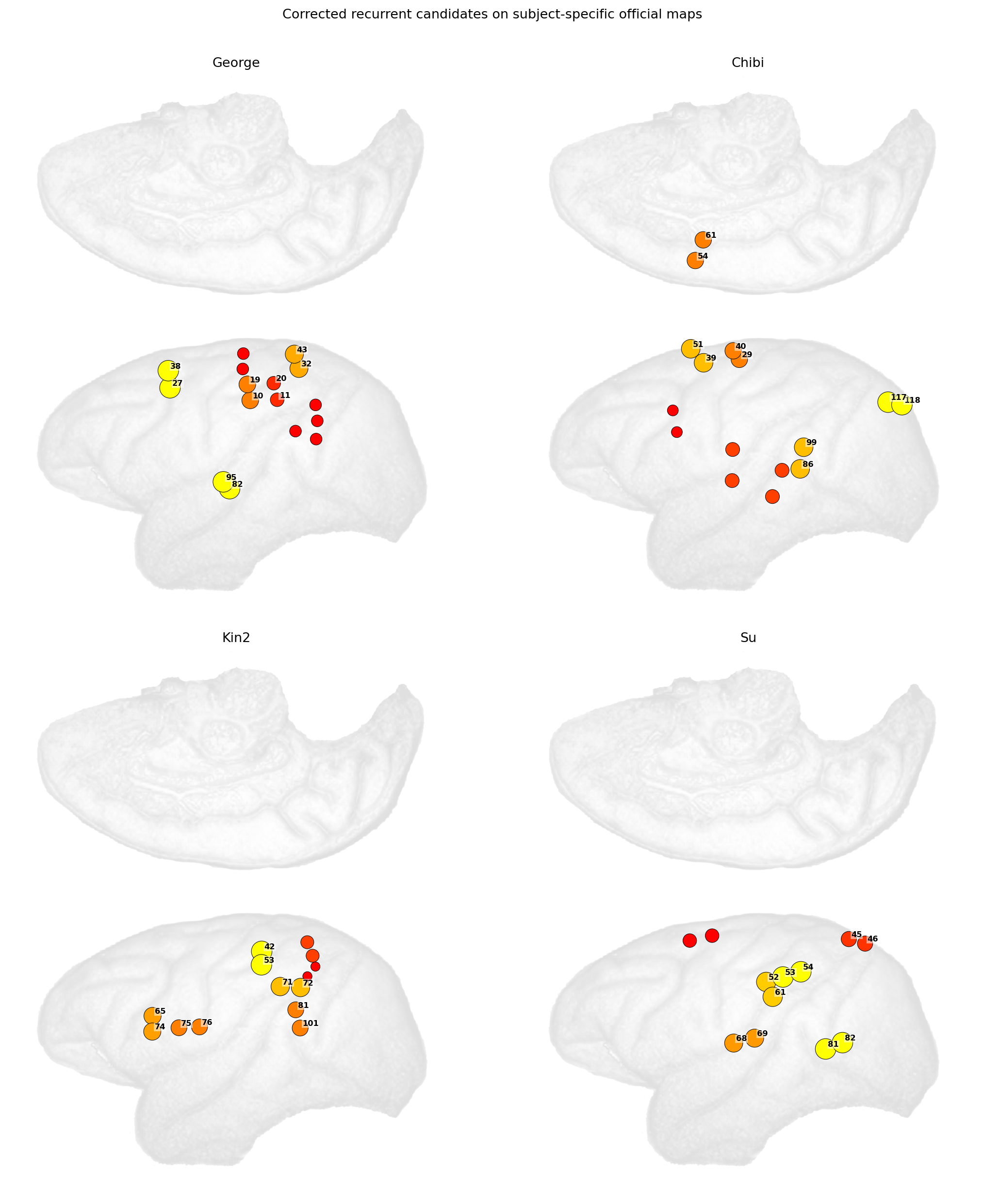}
    \caption{\textbf{Completed subject-specific electrode-layout localization.} Bubble size and color encode balanced same-animal LODO candidate recurrence averaged across $k=3$--5. All panels use the final montage-specific analyses. Selection frequency measures recurrence under the estimator and is not a probability of causal necessity.}
    \label{fig:macaque-official-maps}
\end{figure}

\begin{figure}[t]
    \centering
    \includegraphics[width=0.96\textwidth]{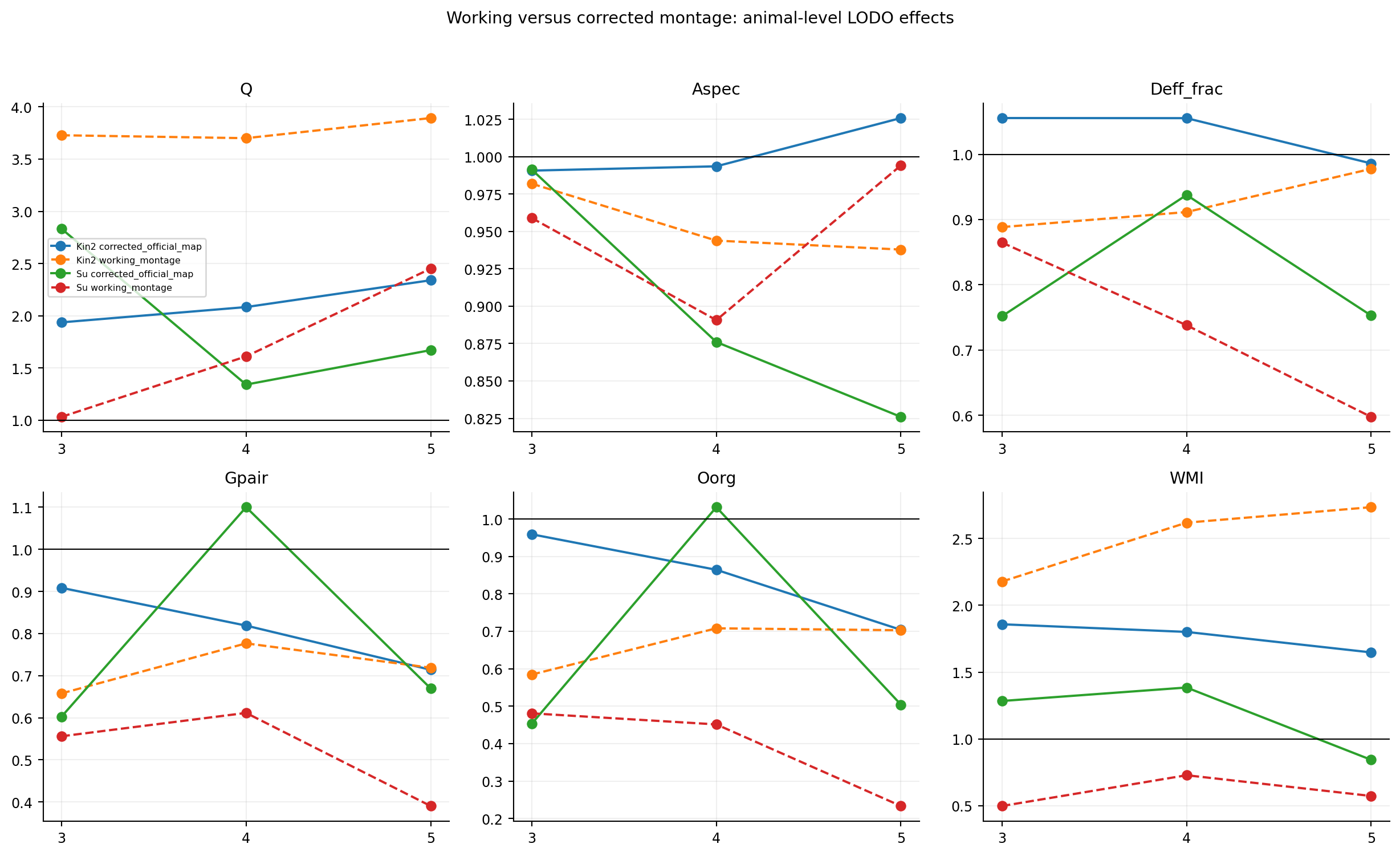}
    \caption{\textbf{Montage sensitivity in the two rematched animals.} Working-montage and corrected geometry-aware LODO ratios are compared by animal and candidate size. The increase in gain and the general reduction in alignment survive rematching, while exact component magnitudes and candidate identities remain montage dependent.}
    \label{fig:macaque-montage}
\end{figure}

\subsection{Recovery trajectories}

Full-state recovery estimates were available for nine days from three animals. $Q$ and $\Cspec$ remained below the preanesthetic awake level during both eyes-closed and eyes-open recovery. Across $k=3$--5, $Q$ ranged from $0.43$ to $0.59$ of awake during eyes-closed recovery and from $0.47$ to $0.55$ during eyes-open recovery. Alignment approached baseline more closely at $k=4$--5, whereas $\Deff/k$ remained modestly below one. $G_{\mathrm{pair}}$ and $\Oorg$ had wide intervals and could lie at or above baseline depending on candidate size. Drug concentration was not matched between induction and recovery, so these component-specific trajectories should not be interpreted as pharmacological hysteresis.

\begin{figure}[t]
    \centering
    \includegraphics[width=0.96\textwidth]{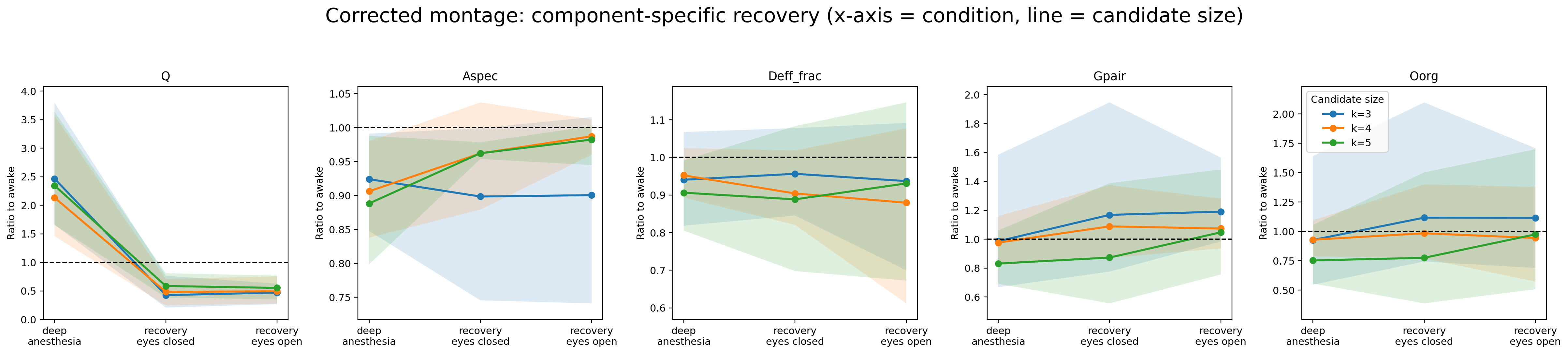}
    \caption{\textbf{Component-specific recovery after corrected rematching.} Ratios to preanesthetic awake data are shown for deep anesthesia, eyes-closed recovery, and eyes-open recovery. Separate lines indicate $k=3,4,5$; intervals are descriptive animal-cluster bootstrap intervals for the nine confirmed recovery days.}
    \label{fig:macaque-recovery}
\end{figure}

\begin{figure}[p]
    \centering
    \includegraphics[width=0.99\textwidth]{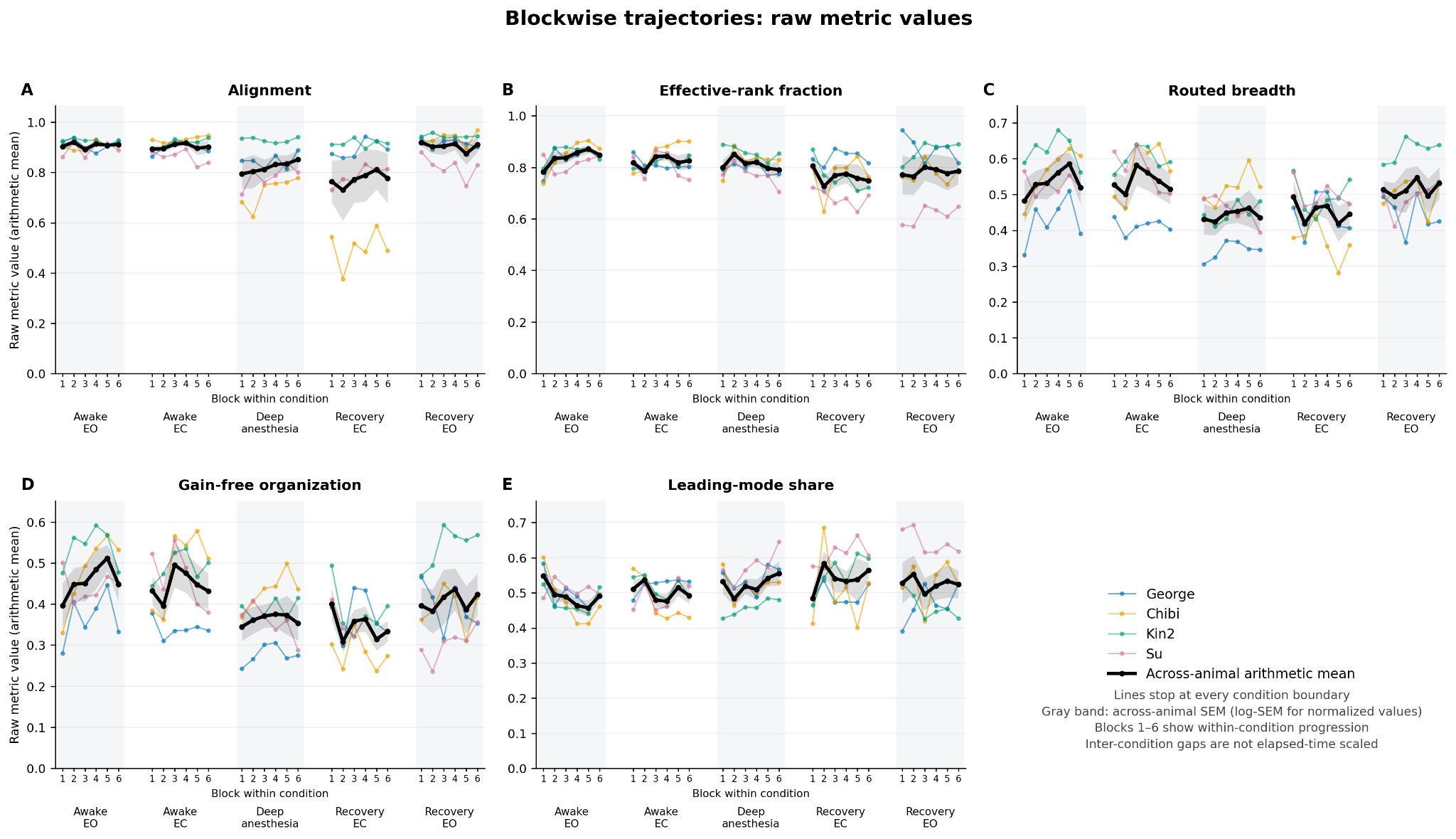}
    \caption{\textbf{Unnormalized blockwise trajectories in the original metric units.} Six equal-duration blocks are shown for each condition. Colored lines show animal-level trajectories, the black line is the across-animal arithmetic mean, and the gray band shows the across-animal standard error. Lines stop at condition boundaries. The corresponding awake-centered normalized display appears in the main text.}
    \label{fig:macaque-blockwise-raw}
\end{figure}

\subsection{Static nonlinear prediction audit}

In the continuous record from Chibi (C2 in the NeuroTycho archive), linear, tanh-residual, and quadratic-residual predictors were compared on held-out temporal segments. The static nonlinear terms did not improve held-out prediction. This negative result was retained to prevent trajectory-conditioned effects from being misrepresented as a general advantage of nonlinear regression.

\begin{figure}[t]
    \centering
    \includegraphics[width=0.88\textwidth]{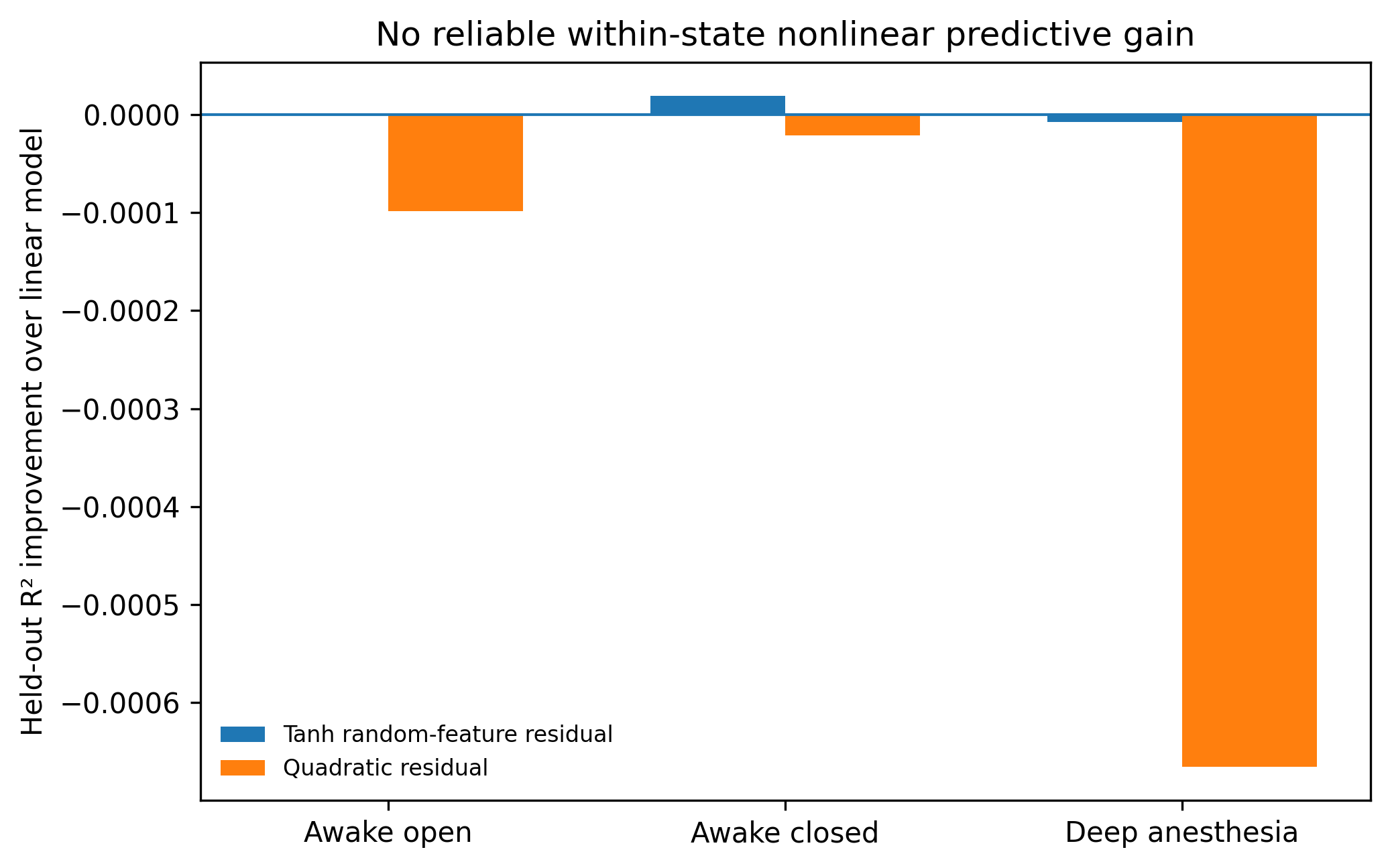}
    \caption{\textbf{Static nonlinear predictive audit in Chibi (C2).} Tanh and quadratic residual terms did not outperform the linear baseline on held-out data.}
    \label{fig:macaque-static-nl}
\end{figure}

\subsection{Exploratory single-animal continuous-induction analysis}

\begin{figure}[p]
    \centering
    \begin{subfigure}[t]{0.96\textwidth}
        \centering
        \includegraphics[width=\textwidth]{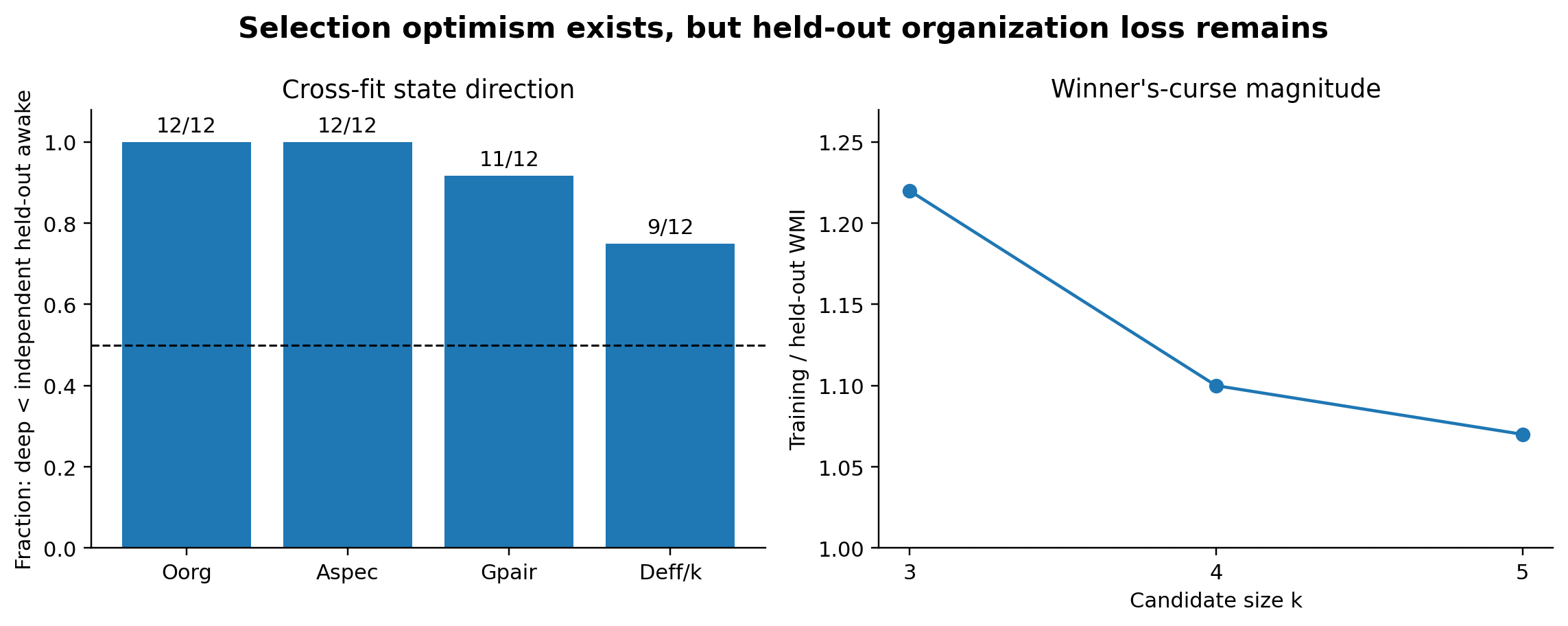}
        \caption{Awake-constrained instantiating-coalition selection audit.}
    \end{subfigure}
    \vspace{0.8em}
    \begin{subfigure}[t]{0.96\textwidth}
        \centering
        \includegraphics[width=\textwidth]{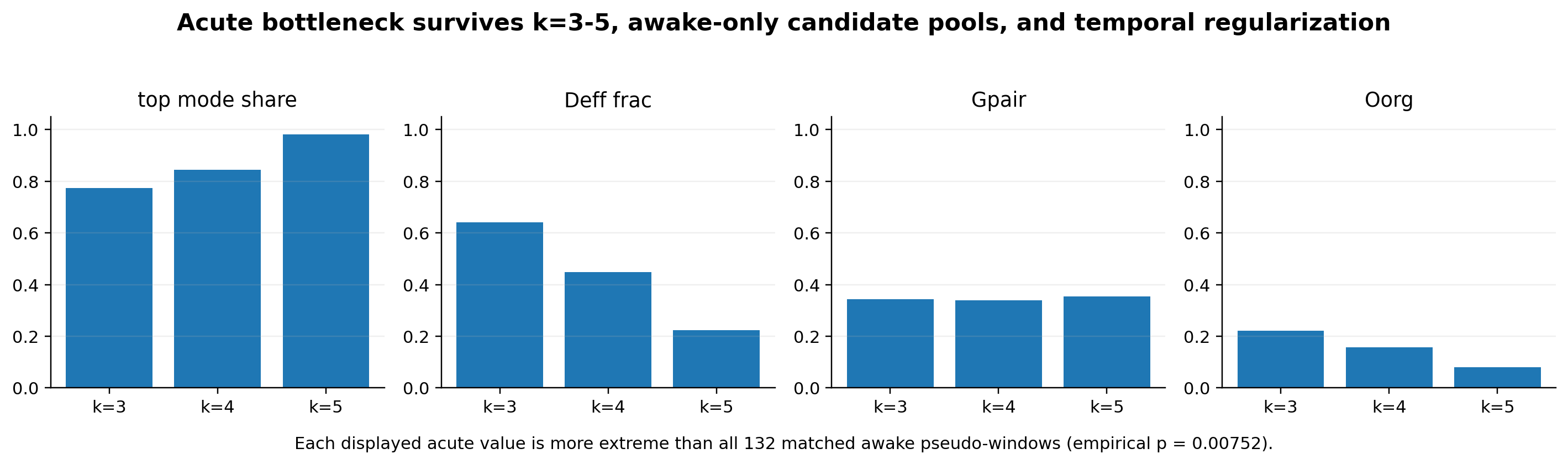}
        \caption{Candidate-size and acute-bottleneck sensitivity.}
    \end{subfigure}
    \caption{\textbf{Selection and size audits for the Chibi nonlinear analysis.} The acute bottleneck survives constrained comparisons, but no result identifies a unique canonical core size.}
    \label{fig:macaque-dynamic-audits}
\end{figure}

This exploratory appendix analysis is retained separately from the main cross-animal trajectory result. Local state-dependent operators were estimated over moving windows in Chibi (C2). Candidate search was constrained to an awake-derived pool, regularized temporally, and compared with 132 awake pseudo-windows. The acute near-rank-one bottleneck occurred near deep-state onset and was robust across $k=3$--5, but the exact maximizing set and the extremity of size-normalized measures depend on the declared candidate pool and size. Additional selection-bias and size audits are shown in \cref{fig:macaque-dynamic-audits}.

\section{Notation summary}
\label{app:notation}

\small
\begin{longtable}{p{0.19\textwidth}p{0.73\textwidth}}
\toprule
Symbol & Meaning \\
\midrule
$S,R$ & Candidate node set and its remainder \\
$A_S,B_S,C_S$ & Candidate internal, boundary-input, and boundary-output operators \\
$L,q$ & Number of past/future samples and required minimum internal shift \\
$\Reach_L,\Observe_L$ & Finite-horizon boundary-reachability and observability matrices \\
$\Hankel_{L,q}$ & Linear shifted boundary-Hankel operator $\Observe_LA_S^q\Reach_L$ \\
$\eta_i,Q$ & Mediation singular values and their sum \\
$\Wc,\Wo$ & Boundary-reachability and observability Gramians; $\Wc$ has the classical controllability interpretation only for independently specified inputs \\
$\Cspec$ & Spectral capacity envelope \\
$\Aspec$ & Realized spectral alignment $Q/\Cspec$ \\
$\Deff$ & Entropy effective rank of the mediation spectrum \\
$G_{\mathrm{pair}}$ & Normalized entropy breadth of routed source--target module energies \\
$\Oorg$ & Gain-free organization $(\Deff/k)G_{\mathrm{pair}}$ \\
WMI & Pragmatic fixed-size scalar $Q(\Deff/k)G_{\mathrm{pair}}$ \\
$\mathcal H^{\mathrm{NL}}$ & Nonlinear past-to-future boundary map for a candidate GMW \\
$\Reach^-,\Observe^+$ & Trajectory-conditioned differential reachability and observability matrices \\
$K_{\epsilon,\mu}$ & Finite-amplitude ensemble secant operator \\
\bottomrule
\end{longtable}
\normalsize
\FloatBarrier
\clearpage
\bibliographystyleapp{plainnat}
\bibliographyapp{references}

\end{document}